\documentclass[preprint,aps,prd,amsmath,amssymb,floatfix]{revtex4}
\pdfoutput=1
\usepackage{graphicx}
\usepackage{color}
\newcommand{\be}{\begin{equation}}
\newcommand{\ee}{\end{equation}}
\newcommand{\bea}{\begin{eqnarray}}
\newcommand{\eea}{\end{eqnarray}}
\newcommand{\beq}{\begin{eqnarray}}
\newcommand{\eeq}{\end{eqnarray}}

\newcommand{\bmp}{\noindent\begin{minipage}{16cm}}
\newcommand{\emp}{\end{minipage}\vskip 7mm} 

\usepackage{graphicx}
\usepackage{dcolumn}
\usepackage{bm}
\usepackage{amsmath}
\usepackage{amsfonts}
\usepackage{bbm}
\usepackage{subfigure}
\usepackage{pxfonts}
\usepackage{slashed}

\usepackage{ulem}

\def\drawbox#1#2{\hrule height#2pt
        \hbox{\vrule width#2pt height#1pt \kern#1pt
              \vrule width#2pt}
              \hrule height#2pt}

\def\Asym#1#2{\vcenter{\vbox{\drawbox{#1}{#2}
              \kern-#2pt 
              \drawbox{#1}{#2}}}}

\begin{document}

\title{ 
{\LARGE {Gravitational {\color{red} T}e{\color{blue}c}{\color{green}h}{\color{yellow}n}{\color{magenta}i}waves}} }
\author{Matti {\sc J\"arvinen}$^{\color{blue}{\varheartsuit}}$}
\email{mjarvine@cp3.sdu.dk}
\author{Chris {\sc Kouvaris}$^{\color{blue}{\spadesuit}}$}
\email{ckouvari@ulb.ac.be}
\author{Francesco {\sc Sannino}$^{\color{blue}{\varheartsuit}}$}
\email{sannino@cp3.sdu.dk} \affiliation{$^{\color{blue}{\varheartsuit}}$ {CP$^\textrm{3}$-Origins, University of Southern Denmark, Campusvej 55, DK-5230 Odense M} \\ $^{\color{blue}{\spadesuit}}${Service de Physique Th\'eorique,  Universit\'e Libre de Bruxelles, 1050 Brussels, Belgium}}

\begin{flushright}
{\it CP$^{\rm \it 3}$-Origins-2009-24}
\end{flushright}

\begin{abstract}
We investigate the production and possible detection of gravitational waves stemming from the electroweak phase transition in the early universe in models of minimal walking technicolor. In particular we discuss the two possible scenarios in which one has only one electroweak phase transition and the case in which the technicolor dynamics allows for multiple phase transitions. 
\end{abstract}

\maketitle

\section{Introduction}

 Recent progress in the understanding of the phase diagram of generic asymptotically free gauge theories \cite{Sannino:2009za,Sannino:2004qp,Dietrich:2005jn,Dietrich:2006cm,Ryttov:2007sr,Ryttov:2007cx} has led to  a renewed interest in this class   of models \cite{Weinberg:1979bn,Susskind:1978ms}. {}For a recent review of the latest developments see \cite{Sannino:2009za}.  {}Explicit examples of technicolor models, not in conflict with electroweak precision tests, have been put forward in \cite{Sannino:2004qp,Dietrich:2005jn,Dietrich:2006cm,Foadi:2007ue,Ryttov:2008xe}. The simplest incarnations of these models are known as (Ultra) Minimal Walking Technicolor models \cite{Sannino:2004qp,Dietrich:2005jn,Dietrich:2006cm,Ryttov:2008xe,Gudnason:2006ug} and indicated in short by MWT and UMT respectively. The principal feature is that the gauge dynamics is such that one achieves (near) conformal dynamics for a small number of flavors and colors. It has been shown that one can construct cold dark matter candidates via either the lightest technibaryon, here termed Technicolor Interacting Massive Particles (TIMP)s \cite{Nussinov:1985xr,Chivukula:1989qb,Barr:1990ca,Gudnason:2006ug,Gudnason:2006yj,Kouvaris:2007iq,Kouvaris:2008hc,Ryttov:2008xe,Foadi:2008qv}, or new heavy leptons naturally associated to the technicolor theory  \cite{Kainulainen:2006wq,Khlopov:2007ic,Khlopov:2008ty}. The TIMP is naturally of asymmetric dark matter type \cite{Nussinov:1985xr,Chivukula:1989qb,Barr:1990ca}, meaning that its relic density does not have a thermal origin. Within the (U)MWT models such a relic density has been estimated in \cite{Gudnason:2006ug,Gudnason:2006yj,Ryttov:2008xe}. Weak isotriplet TIMPs (iTIMP)s have been shown to be interesting candidates of dark matter in 
 \cite{Frandsen:2009mi}. 

Another interesting cosmological arena is the temperature driven electroweak phase transition within technicolor theories \cite{Cline:2008hr,Jarvinen:2009wr,Jarvinen:2009pk,Kikukawa:2007zk}.  We have investigated in much detail this phase transition for the MWT and UMT models using the low effective Lagrangian approach. We discovered that there is a sizable region of the low energy effective theories' parameters yielding a sufficiently strong first order electroweak phase transition to drive, in principle, electroweak baryogenesis. We have also discovered quite a rich phase diagram in the case of the UMT model \cite{Jarvinen:2009pk} and more generally whenever several underlying matter representations are simultaneously present in the technicolor dynamics \cite{Jarvinen:2009wr}. An interesting problem is if such a transition is observable via detecting the cosmological gravitational waves (GW)s produced at the transition itself. We will describe the topic of GWs in more detail in the next section. Whether or not these waves are observable depends on the strength of the electroweak phase transition. We will investigate this issue using the two concrete models discussed above.  To be able to study the production of GWs, we need to use a slightly improved treatment of the phase transition compared to our earlier work. We confirm the results of \cite{Cline:2008hr,Jarvinen:2009pk}, and find that the MWT model Lagrangian can support a sufficiently strong electroweak phase transition leading to an observable signal at BBO \cite{Harry:2006fi}. An interesting feature of the UMT model is the presence of multiple electroweak phase transitions arising at different temperatures. This  is due to the interplay between two distinct chiral phase transitions, one directly responsible for the electroweak symmetry breaking and the other decoupled from the standard model (SM). However we find that the planned experiments searching for GWs will have hard times discovering the signal originating from this model and more sensitive ones are needed. 
 
 It is, however, possible to increase the strength of the first order phase transition by considering {\it partially} gauged technicolor models \cite{Dietrich:2005jn,Dietrich:2006cm}. They have, by construction, a large number of techniflavors but only two of them are gauged under the electroweak symmetry. This choice reduces the contribution to the electroweak precision parameters, while the large number of techniflavors enhances the strength of the first order phase transition. 
The nonrenormalizable axial anomalous contributions to the effective low energy potential is partially responsible for increasing the strength of the transition when increasing the number of techniflavors.  We will, however, investigate the spectrum of gravitational waves associated to the electroweak phase transition stemming from generic models of partially gauged technicolor elsewhere. For other simple models which predict potentially strong GWs, see, for example, \cite{Delaunay:2007wb,Kehayias:2009tn}.  

Summarizing we investigate in detail the MWT and UMT models, at the effective Lagrangian level, and show that the MWT can lead to detectable gravitational waves while the UMT cannot.  

\section{Gravitational Waves Production Setup }

In this section we lay the basics of the GW production from strong first order phase transitions and we present the relevant parameters of our theories needed for the calculation of the gravitational signal.

First of all, let us review how a first order phase transition takes place in the early universe and why it can produce GWs. In a first order phase transition there are two distinct minima separated by a potential barrier. The phase transition can be thought to start taking place at the moment where the two vacua are at the same energy level. Immediately after, the true-vacuum state to be lowers its potential level compared to the one of the false vacuum and therefore despite the existence of the barrier, quantum mechanically, there is a finite probability for the system to pass from the false to the true vacuum. The phase transition occurs through nucleation of bubbles of the true vacuum~\cite{Coleman:1977py}. (For a nice exposition see~\cite{Kolb:1990vq}.) The nucleation is possible due to quantum tunneling and thermal fluctuations. The bubbles of the true vacuum expand until they cover the whole space, which means that the  phase transition has been concluded. For the production of GWs a quadrupole moment is required, and since the bubbles are spherical, and therefore have no quadrupole, it seems at first sight that no production can take place. However, there are at least two different ways of producing GWs from bubbles. The first one is when bubbles collide. Apparently in this case the spherical symmetry is destroyed and GWs are produced. The second source of GW production is due to turbulence of the plasma because of the bubble's motion.

For a given theory, there are basically two parameters that determine the signal of the GWs produced \cite{Grojean:2006bp,Kamionkowski:1993fg,Kosowsky:2001xp}. The first one $\alpha$ is defined as the ratio of the latent heat $\epsilon$ of the phase transition at the bubble nucleation temperature over the energy density of the false vacuum. Practically the latent heat is the energy released as the system tunnels from the false to the true vacuum. It is given by
\bea \epsilon = -\Delta V -T \Delta s = - \Delta V +T \frac{\partial V}{\partial T}  \ , 
\eea where $V$ is the potential.
 Since we are interested in theories where the phase transition takes place around the electroweak scale ($\sim 250$ GeV), the energy density is dominated by the radiation part. The parameter $\alpha$ practically measures how strong the phase transition is. As we shall discuss later on, large $\alpha$, i.e., strong first order phase transition, leads to enhanced amplitude for the GW and therefore better detectability. From this point of view, theories with strong phase transitions are more interesting.

If roughly speaking $\alpha$ affects the amplitude of the GWs, the second {\it model dependent} parameter $\beta$, determines the characteristic frequency. This is because $\beta^{-1}$ corresponds to the rate of change of the nucleation probability and therefore has units of inverse time. This means that $\beta^{-1}$ is approximately the duration of the phase transition, and provided we know the velocity of the bubble expansion, it determines the size of the bubble (having ignored the initial size which is negligible compared to the final). Let us see this explicitly. The bubble nucleation rate  at nonzero temperature is given by
\bea \Gamma  \simeq T^4 e^{-S_E}, \eea
where $S_E=S_3/T$ and
\bea S_3= \int dr 4\pi r^2 \left [ \frac{1}{2} \left ( \frac{d\phi}{dr} \right )^2 + V(\phi, T) \right ] \label {s3} \eea is the Euclidean 3-dimensional action. $\phi$ is  the bubble profile. We are looking for a least action solution which has an $O(3)$ symmetry. The equation of motion exhibiting manifestly this symmetry and dictated by the minimization of the Euclidean action is
\bea \frac{d^2\phi}{dr^2}+\frac{2}{r}\frac{d\phi}{dr}-\frac{dV(\phi,T)}{d\phi}=0, \label{eqmotion} \eea
where $r$ is the radial coordinate. The boundary conditions are $d\phi (r=0)/dr=0$, i.e.,  we require the solution to be smooth at 
the center of the created bubble, and $\phi(r=\infty)=0$, meaning that far away from the bubble the system is still in the false vacuum. If one imagines $\phi$ to be the position of a particle and $r$ to be the time, the above equation corresponds to the equation of motion of a particle within a potential $-V(\phi)$ with unit mass and a Stokes type of drag force proportional to the velocity given by the second term of the equation above. In general, the bounce solution cannot be found analytically due to the complexity of the equation. However within the thin wall approximation a closed solution has been found. The thin wall approximation is valid when the difference in the height of the two minima is small compared to the barrier. Let us call $\phi_e$ the ``escape point'', i.e., the value of $\phi(r=0)$ of the solution of Eq.~(\ref{eqmotion}). In the particular case where the two minima are almost degenerate, $\phi_e$ should be very close to the true minimum 
to reduce the work done against the drag force: 
The particle starts from a nearly flat point in the potential and therefore it would take some time ($r$ in this particular case) to build up its velocity, and then go downhill fast in order to come at rest again at the false minimum.
Practically, this means that within this approximation, the friction term of Eq.~(\ref{eqmotion}) 
can be safely ignored and an analytical result can be obtained. It also means that the change from one minimum to the other happens ``fast'' and the profile of $\phi$ is quite sharp. This justifies the name ``thin wall.'' Generally, if one is not sure whether or not the thin wall approximation is valid, a numerical solution of Eq.~(\ref{eqmotion}) is needed. The standard way of finding this numerical solution (which is the one we also used in order to get our results) is by guessing the value of the $\phi_e$. We know that $\phi_e$ should be between the two minima. If our guessed value of $\phi_e$ is closer to the true minimum that in reality, solving Eq.~(\ref{eqmotion}) with the boundary conditions $\phi(r=0)=\phi_e^{\text{trial}}$ and $d\phi(r=0)/dr=0$ will ``overshoot'' the solution, meaning that after some $r$ the solution will start going to $-\infty$. On the other hand, if $\phi_e^{\text{trial}}$ is closer to the false minimum than the actual $\phi_e$, the trial solution will ``undershoot'' the real solution meaning that it will never reach the false vacuum.

Having introduced the Euclidean action, $\beta$ is defined as the time derivative of the action $-dS_E/dt$, evaluated at the nucleation temperature that we shall introduce shortly. In the early universe the expansion parameter $a\sim T^{-1}$ and therefore the Hubble parameter $H=(1/a)da/dt=-(1/T)dT/dt$. 
Consequently \bea \frac{\beta}{H}=T\frac{dS_E}{dT}=T\frac{d(S_3/T)}{dT}. \eea It is understood that everything is evaluated at the nucleation temperature. This temperature is defined as the temperature where the rate of bubble nucleation per Hubble volume and time is approximately one. This means
\bea \Gamma \simeq H^4 \rightarrow T\ln\frac{T}{m_{Pl}}\simeq -\frac{S_3}{4}, \eea where $m_{Pl}$ is the Planck mass and we used $H \simeq T^2/m_{Pl}$. The above equation gives the bubble nucleation temperature. 
We shall denote this temperature by $T_*$ and the value of the Hubble parameter at nucleation by $H_*$ below.  
Recall that $S_3$ is known once we have found the bounce solution, and substitute it in Eq.~(\ref{s3}).

The parameters $\alpha$ and $\beta$ are the essential input parameters we need from the specific model under investigation. The strength and the frequency of the gravitational signal produced by the first order phase transition are encoded in these two parameters. Let us review the basic arguments of how GWs are produced due to bubble collisions in a more quantitative way, following the scaling argument presented in~\cite{Grojean:2006bp}. GWs are produced through quadrupole (or higher moment) emission. For the quadrupole, the GW power is $P=(G/5)\langle (\dddot{Q_{ij}^T})^2 \rangle$, where $G$ is the Newton constant and $Q_{ij}$ is the quadrupole moment of the transverse and traceless part of the energy-momentum tensor. Note the dependence of the power on the triple derivative of the quadrupole with respect to time, something which is also true in electromagnetism. The quadrupole moment has dimensions of mass times distance squared and therefore dimension analysis dictates that the triple derivative would have the units of kinetic energy over time. Not all the energy gained from tunneling from the false to the true vacuum is in the form of kinetic energy. If $k$ is the fraction of the latent heat in the form of kinetic energy (the rest being heat), $E_{\text{kin}}\sim k \alpha \rho_{\text{rad}}(v_b/ \beta)^3$. As we have already mentioned the latent heat is $\alpha \rho_{\text{rad}}$, and we also have multiplied by the volume of the bubble $\sim (v_b/\beta)^3$ 
where $v_b$ is the velocity of the bubble walls. 
From Friedmann's equation we know that $ H_*^2 \sim G \rho_{\text{crit}}$, and therefore we can trade $G$ for $H_*$. In addition, $\rho_{\text{crit}}=(1+\alpha)\rho_{\text{rad}}$. Using all of the above and keeping in mind that $E_{GW}=P/\beta$, we get that $\Omega_{GW} \sim (H_*/\beta)^2k^2 \alpha^2 v_b^3/(1+\alpha)^2$. This is as good as dimension analysis can get us. 

The GW production due to bubble collisions was first studied in~\cite{Kosowsky:1991ua,Kosowsky:1992rz,Kosowsky:1992vn,Kamionkowski:1993fg}. These calculations were based on numerical simulations of bubbles colliding using the so-called envelope approximation, which consists of considering only the nonoverlapping regions of the collided bubbles as sources of GW production. In such case~\cite{Kamionkowski:1993fg}
\bea \Omega_{\text{coll}}h^2 \simeq 1.1 \times 10^{-6}k^2 \left ( \frac{H_*}{\beta} \right )^2 \left (\frac{\alpha}{\alpha+1} \right )^2 \frac{v_b^3}{0.24+v_b^3} \left (\frac{100}{g_*} \right)^{1/3} \label{col1} \ , \eea 
where $g_*$ is the number of relativistic degrees of freedom at the nucleation temperature.
Assuming a detonation, the bubble wall velocity is given approximately by~\cite{Steinhardt:1981ct}
\bea v_b(\alpha)=\frac{1/\sqrt{3}+\sqrt{\alpha^2+2\alpha/3}}{1+\alpha}. \eea In addition~\cite{Kamionkowski:1993fg} \bea k(\alpha)\simeq \frac{1}{1+0.715\alpha}\left [0.715\alpha+\frac{4}{27}\sqrt{\frac{3\alpha}{2}} \right ]. \eea The peak frequency is
\bea f_{\text{coll}}\simeq 5.2 \times 10^{-6} \left (\frac{\beta}{H_*} \right ) \left (\frac{T_*}{100 \text{GeV}} \right ) \left (\frac{g_*}{100} \right )^{1/6}\text{Hz}. \eea
Analysis of two-bubble collisions suggests that the spectrum rises as $f^{2.8}$ and $f^{-1.8}$, below and above the peak frequency, respectively \cite{Kosowsky:1991ua}.

The subject of the GW production from first order phase transitions is still a field of active research and of continuous developments. The authors of~\cite{Caprini:2007xq}
developed a different modeling of the problem. Instead of performing numerical simulations of colliding bubbles, they considered the bubble wall velocity as a random variable. Although in this approach the collisions are not formulated in a deterministic way, the advantage is that the envelope approximation in this case is not implemented. The spectrum is
\bea \Omega_{\text{coll}}'h^2\simeq 9.8 \times 10^{-8} v_f^4\frac{(1-s^3)^2}{(1-s^2v_f^2)^4} \left (\frac{H_*}{\beta} \right )^2 \left ( \frac{100}{g_*} \right )^{1/3} \label{col3} \ , \eea 
where  
$v_f=(v_b-1/\sqrt{3})/(1-v_b/\sqrt{3})$, and $s=1/(v_b\sqrt{3})$. The peak frequency is
\bea f_{\text{coll}}' \simeq 1.12 \times 10^{-5} \frac{\beta}{H_*} \frac{T_*}{100 \text{GeV}} \left (\frac{g_*}{100} \right )^{1/6} \frac{1}{v_b} \text{Hz}. \eea
Away from the peak frequency, the spectrum is multiplied by a factor $2.5 f_r^3/(1+0.5f_r^2+f_r^{4.8})$, where $f_r=f/(0.87 f_{\text{coll}}')$. Finally, new numerical simulations have been done recently with multicolliding bubbles~\cite{Huber:2008hg} suggesting that the GW spectrum decreases like $f^{-1}$ 
rather 
than $f^{-1.8}$. This calculation gives
\bea \Omega_{\text{coll}}''h^2 \simeq 1.84 \times 10^{-6}k^2 \left (\frac{ H_*}{\beta} \right )^2 \left (\frac{\alpha}{\alpha+1} \right )^2 \frac{v_b^3}{0.42+v_b^2} \left (\frac{100}{g_*} \right)^{1/3}, \label{col2} \eea with peak frequency
\bea f_{\text{coll}}''=1.65 \times 10^{-5} \left ( \frac{\beta}{H_*} \right ) \left ( \frac{T_*}{100 \text{GeV}} \right ) \left ( \frac{g_*}{100} \right )^{1/6} \frac{0.62}{1.8-0.1v_b+v_b^2} \text{Hz}. \eea The spectrum (according to this calculation) rises as $f^3$ for frequencies below the $f''$ and falls off as $f^{-1}$ for frequencies larger than $f''$.

As we have mentioned, the second source of GWs during a first order phase transition can be turbulence. When bubbles collide, the plasma is stirred up and develops the characteristics of a fully developed turbulence. This means that a cascade of eddies is created in the plasma. Large eddies (of the size of the system or the stirring source) are formed and after a few revolutions, they break down to smaller eddies until their size becomes equal to the damping scale. Fluid (nonrelativistic) turbulence has been found experimentally to agree with Kolmogorov's stochastic description. This description has been implemented in calculations of GW production due to turbulence~\cite{Kosowsky:2001xp,Dolgov:2002ra,Caprini:2006jb,Gogoberidze:2007an}, although the fluid in this case is relativistic. Although in all these calculations, the Kolmogorov spectrum is used in order to model turbulence, there is a sort of different philosophy between for example~\cite{Dolgov:2002ra}, and~\cite{Caprini:2006jb}. In the former, GWs inherit directly the frequency of the eddies, while in the latter GWs inherit the wave number of the eddies. It is easy to see that the two approaches are not equivalent. The characteristic frequency of the eddies is $\omega_l=v_s/l$ (with $v_s$ being the velocity of the fluid in the eddy and $l$ the characteristic length of it). In the first approach, the GWs peak at the frequency of the largest eddy which is $\omega_L=v_s/L$ (with $L$ being the size of the stirring source). On the other hand, the wave number of the stirring source is $k\sim 1/L$.  If the GWs inherit the wave number instead of the frequency of the eddies (as in the second approach), the GW spectrum (because its dispersion relation is $\omega= \vert k \vert$) peaks at frequencies $\omega=1/L \neq v_s/L=\omega_L$ (since $v_s \neq 1$). Here we follow the approach presented in \cite{Caprini:2006rd} for the GW production due to first order phase transitions in the early universe.  In this framework the GWs inherit the wave number spectrum.  However, we have also checked that very similar results are obtained when adopting the framework discussed in \cite{Gogoberidze:2007an,Kahniashvili:2008pf}. The GW density  for $v_s<0.5$ is
\begin{displaymath}
\Omega_{\text{turb}}h^2=6.7 \times 10^{-6}v_s^4 v_b^2 \left ( \frac{H_*}{\beta} \right)^2 \left ( \frac{100}{g_*} \right )^{1/3} \left\{ \begin{array}{ll}
\left (\frac{1}{4v_s^2} \right ) \left (\frac{f}{f_p} \right )^3, & f<2v_sf_p \\
\left ( \frac{f}{f_p} \right ), & 2v_sf_p<f<f_p \\
\left (\frac{f}{f_p} \right )^{-8/3}, & f_p<f
\end{array} \right.,
\end{displaymath}
and for $v_s>0.5$ is

\begin{displaymath}
\Omega_{\text{turb}}h^2=6.7 \times 10^{-6}v_s^4 v_b^2 \left ( \frac{H_*}{\beta} \right)^2 \left ( \frac{100}{g_*} \right )^{1/3} \left (\frac{1}{4v_s^2} \right ) \left\{ \begin{array}{ll}
\left (\frac{f}{f_p} \right )^3, & f<f_p \\
\left ( \frac{f}{f_p} \right )^{-2}, & f_p<f<8v_s^3f_p \\
4v_s^2\left (\frac{f}{f_p} \right )^{-8/3}, & 8v_s^3f_p<f
\end{array} \right.,
\end{displaymath}
where \bea v_s \simeq \sqrt{\frac{k\alpha}{4/3+k\alpha}}. \eea
The peak frequency $f_p$ is~\cite{Caprini:2006rd}
\bea f_{\text{turb}}\simeq 8 \times 10^{-6} \frac{1}{v_b} \left (\frac{\beta}{H_*} \right ) \left ( \frac{T_*}{100~\text{GeV}} \right )\left ( \frac{g_*}{100} \right )^{1/6}~\text{Hz}. \eea

\section{Gravitational Waves from Minimal Walking Technicolor}

The new dynamical sector we consider, which underlies the Higgs
mechanism, is an SU(2) technicolor gauge theory with two adjoint
technifermions \cite{Sannino:2004qp}. The two adjoint fermions may be
written as \beq Q_L^a=\left(\begin{array}{c} U^{a} \\D^{a}
\end{array}\right)_L , \qquad U_R^a \ , \quad D_R^a \ ,  \qquad
a=1,2,3 \ ,\eeq with $a$ being the adjoint color index of SU(2). The
left-handed fields are arranged in three doublets of the SU(2)$_L$
weak interactions in the standard fashion. The condensate is $\langle
\bar{U}U + \bar{D}D \rangle$ which correctly breaks the electroweak
symmetry. The model as described so far suffers from the Witten
topological anomaly \cite{Witten:1982fp}. However, this can easily be
addressed by adding a new weakly charged fermionic doublet which is a
technicolor singlet \cite{Dietrich:2005jn}.

In \cite{Foadi:2007ue} we constructed the effective theory for MWT
including composite scalars and vector bosons, their
self-interactions, and their interactions with the electroweak gauge
fields and the SM fermions. We have also used the Weinberg  modified
sum rules \cite{Appelquist:1998xf} to constrain the low
energy effective theory. This extension of the SM was thereby shown
to pass the electroweak precision tests. Near the finite
temperature phase transition the relevant degrees of freedom are the
scalars and hence we will not consider the vector spectrum or that of
the composite fermions.


The relevant effective theory for the Higgs sector at the electroweak
scale consists, in our model, of a composite Higgs and its
pseudoscalar partner, as well as nine pseudoscalar Goldstone bosons
and their scalar partners. These can be assembled in the matrix
\begin{eqnarray}
M = \left[\frac{\sigma+i{\Theta}}{2} + \sqrt{2}(i\Pi^a+\widetilde{\Pi}^a)\,X^a\right]E \ ,
\label{M}
\end{eqnarray}
which transforms under the full $SU(4)$ group according to
\begin{eqnarray}
M\rightarrow uMu^T \ , \qquad {\rm with} \qquad u\in {\rm SU(4)} \ .
\end{eqnarray}
The $X^a$'s, $a=1,\ldots,9$, are the generators of the $SU(4)$
group which do not leave  the vacuum expectation value (VEV) of $M$
invariant.

The electroweak subgroup can be embedded in SU(4), as explained in detail in \cite{Appelquist:1999dq}.
The new Higgs Lagrangian is
\begin{eqnarray}
{\cal L}_{\rm Higgs} &=& \frac{1}{2}{\rm Tr}\left[D_{\mu}M D^{\mu}M^{\dagger}\right] - {\cal V}(M) + {\cal L}_{\rm ETC} \ ,
\end{eqnarray}
where the potential reads
\begin{eqnarray} \label{Vdef}
{\cal V}(M) & = & - \frac{m^2}{2}{\rm Tr}[MM^{\dagger}] +\frac{\lambda}{4} {\rm Tr}\left[MM^{\dagger} \right]^2
+ \lambda^\prime {\rm Tr}\left[M M^{\dagger} M M^{\dagger}\right] \nonumber \\
& - & 2\lambda^{\prime\prime} \left[{\rm Det}(M) + {\rm Det}(M^\dagger)\right] \ ,
\end{eqnarray}
and ${\cal L}_{\rm ETC}$ contains all terms which are generated by the ETC interactions, and not by the chiral symmetry breaking sector.

We explicitly break the SU(4) symmetry in order to provide mass to the Goldstone bosons which are not eaten by the weak gauge bosons.
Assuming parity invariance,
\begin{eqnarray} \label{VETCdef}
{\cal L}_{\rm ETC} = \frac{m_{\rm ETC}^2}{4}\ {\rm Tr}\left[M B M^\dagger B + M M^\dagger \right] + \cdots \ ,
\end{eqnarray}
where the ellipses represent possible higher dimensional operators, and $B$ is a constant matrix \cite{Foadi:2007ue} that commutes with the SU(2)$_{\rm L}\times$SU(2)$_{\rm R}\times$U(1)$_{\rm V}$ generators.

The potential ${\cal V}(M)$ is SU(4) invariant. It produces a VEV
which parametrizes the techniquark condensate, and spontaneously
breaks SU(4) to SO(4). In terms of the model parameters the VEV is
\begin{eqnarray}
v^2=\langle \sigma \rangle^2 = \frac{m^2}{\lambda + \lambda^\prime - \lambda^{\prime\prime} } \ ,
\label{VEV}
\end{eqnarray}
while the Higgs mass is
\begin{eqnarray}
M_H^2 = 2\ m^2 \ .
\end{eqnarray}
The linear combination $\lambda + \lambda^{\prime} -
\lambda^{\prime\prime}$ corresponds to the Higgs self-coupling in
the SM. The three pseudoscalar mesons $\Pi^\pm$, $\Pi^0$ correspond
to the three massless Goldstone bosons which are absorbed by the
longitudinal degrees of freedom of the $W^\pm$ and $Z$ boson. The
remaining six uneaten Goldstone bosons are technibaryons, and all
acquire tree-level degenerate masses through (not yet specified) ETC interactions \footnote{The Goldstone bosons also receive a mass contribution due to the coupling to the electroweak sector which may be sizeable, see \cite{Dietrich:2009ix}.}:
\begin{eqnarray}
M_{\Pi_{UU}}^2 = M_{\Pi_{UD}}^2 = M_{\Pi_{DD}}^2 = m_{\rm ETC}^2  \ .
\end{eqnarray}
The remaining scalar and pseudoscalar masses are
\begin{eqnarray}
M_{\Theta}^2 & = & 4 v^2 \lambda^{\prime\prime} \nonumber \\
M_{A^\pm}^2 = M_{A^0}^2 & = & 2 v^2 \left(\lambda^{\prime}+\lambda^{\prime\prime}\right)
\end{eqnarray}
for the technimesons, and
\begin{eqnarray}
M_{\widetilde{\Pi}_{UU}}^2 = M_{\widetilde{\Pi}_{UD}}^2 = M_{\widetilde{\Pi}_{DD}}^2 =
m_{\rm ETC}^2 + 2 v^2 \left(\lambda^{\prime} + \lambda^{\prime\prime }\right) \ ,
\end{eqnarray}
for the technibaryons.
Ref.\ \cite{Hong:2004td} provides further insight into some of these
mass relations.

\subsection{Effective potential for MWT}

The electroweak phase transition is studied by using the effective potential method. We include temperature dependent corrections of the effective potential up to one-loop level and ring resummation following Arnold \& Espinosa \cite{Arnold:1992rz}. We follow otherwise the conventions of \cite{Cline:2008hr} but use a slightly different method for estimating the temperature dependent one-loop correction, which involves a combination of the high temperature and low temperature asymptotic series \cite{Cline:1996mga}  (see Appendix~\ref{app:effpot} for details). This improves the potential at low temperatures. This is necessary here since strong GW production requires strong first order phase transitions, which typically means that the critical temperature is much smaller than the electroweak scale. Moreover, we use the actual nucleation temperature $T_*$, which can be considerably smaller than  the critical temperature $T_c$ (where the symmetric phase and broken phase vacua are exactly degenerate) for strong first order transitions.

\begin{figure}[ht]
 {\includegraphics[height=5.1cm,width=5.32cm]{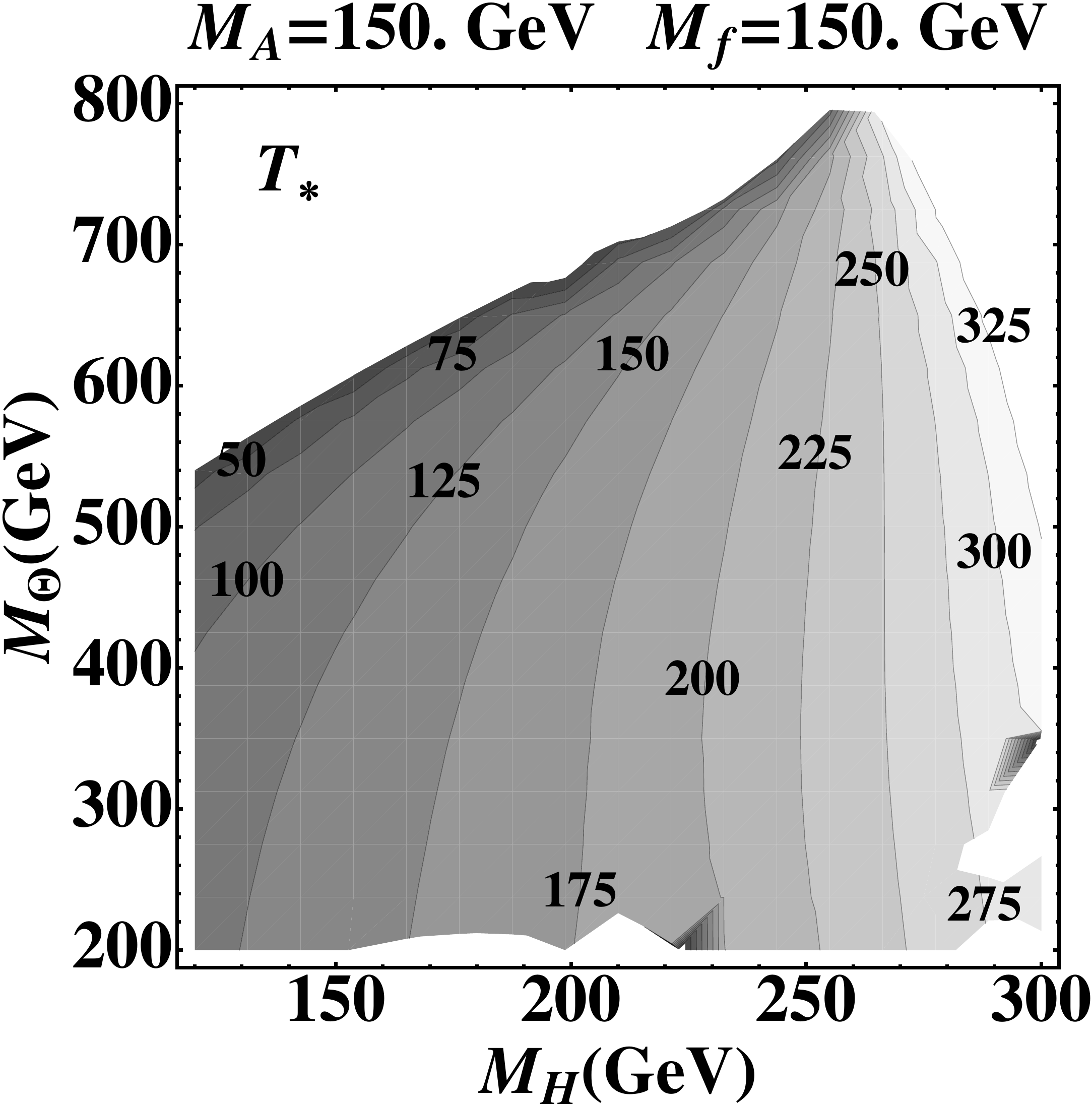}\hspace{0.1cm}\includegraphics[height=5.1cm,width=5.32cm]{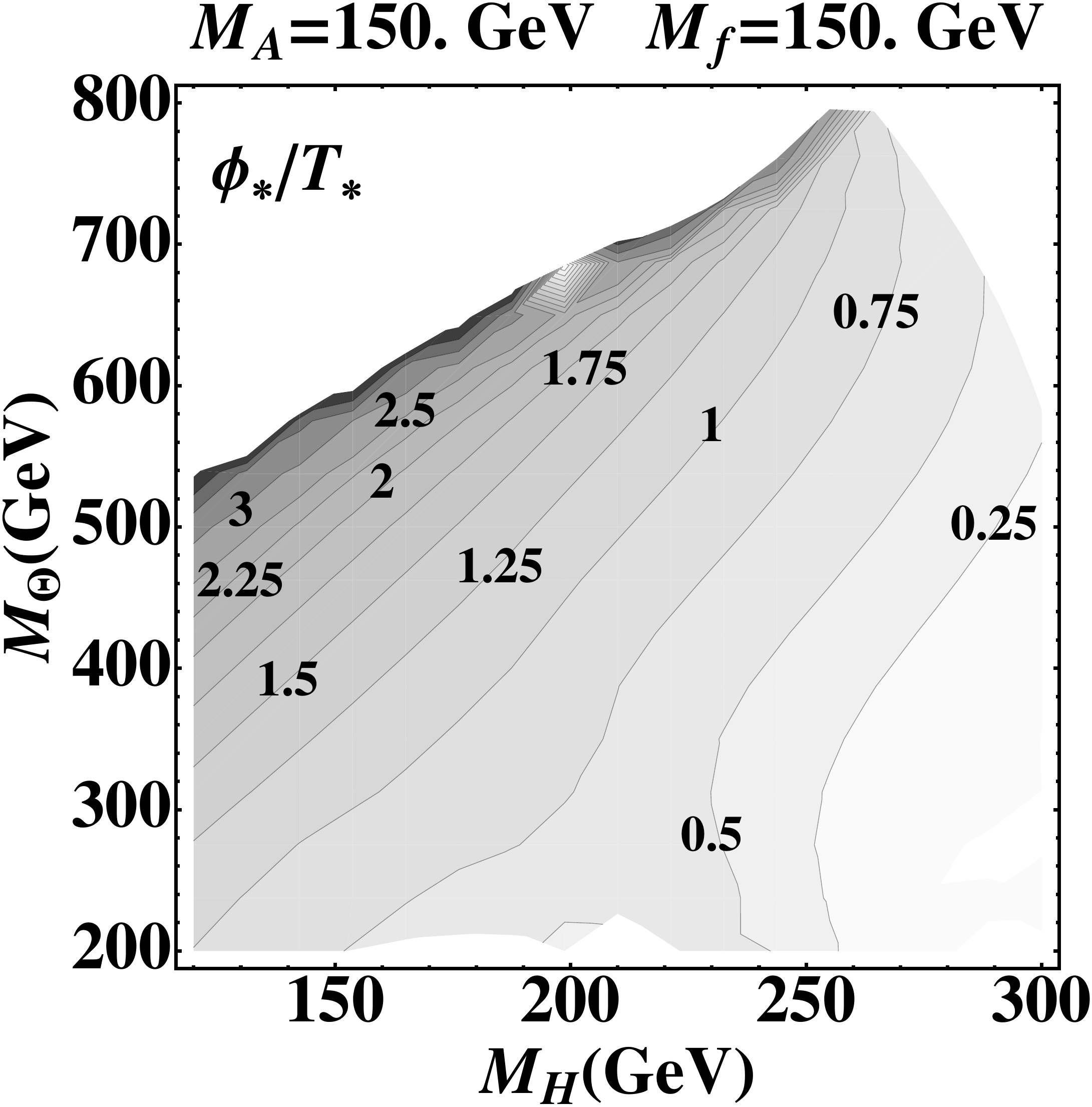}

\includegraphics[height=5.1cm,width=5.32cm]{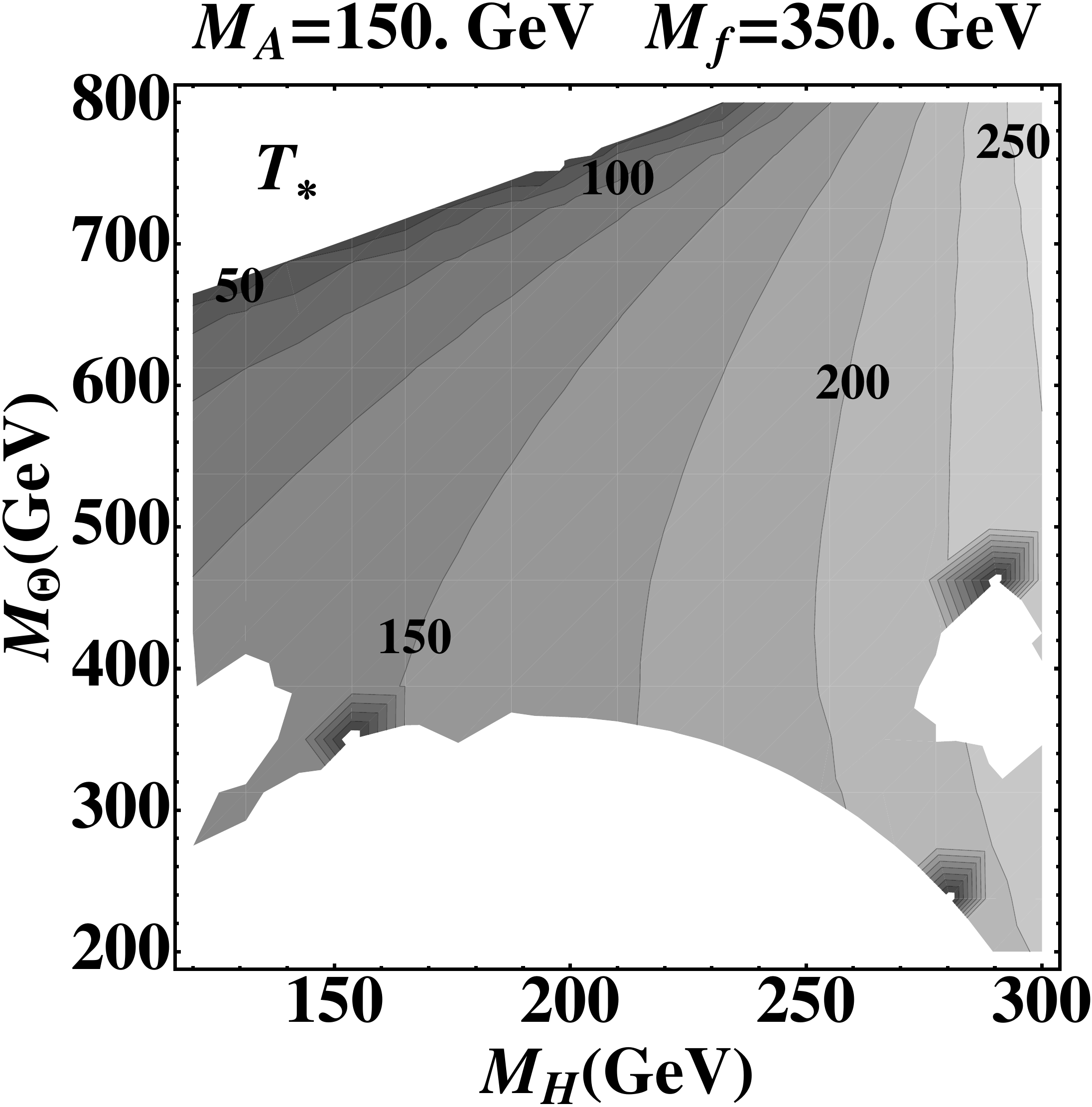}\hspace{0.1cm}\includegraphics[height=5.1cm,width=5.32cm]{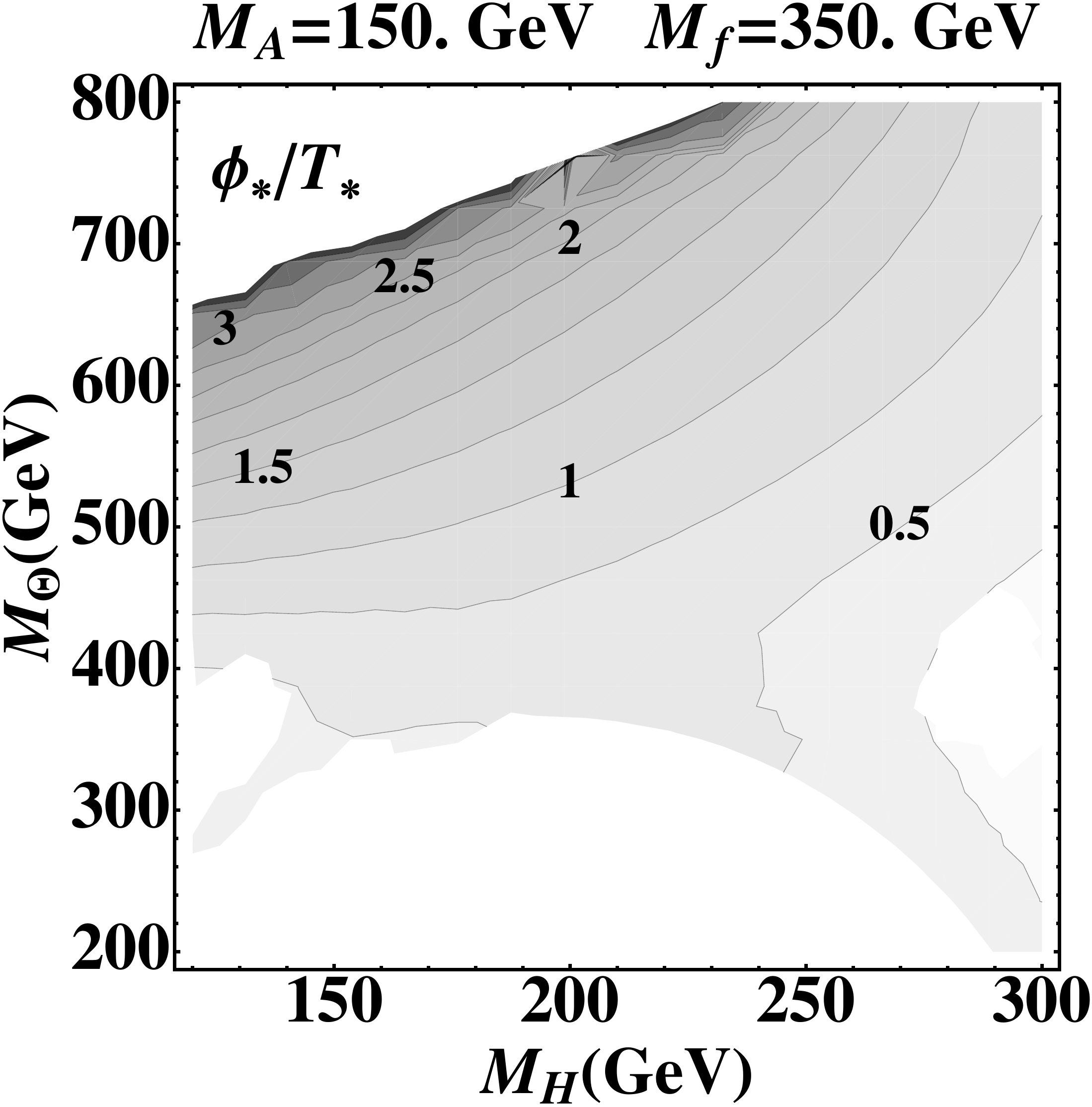}

\includegraphics[height=5.1cm,width=5.32cm]{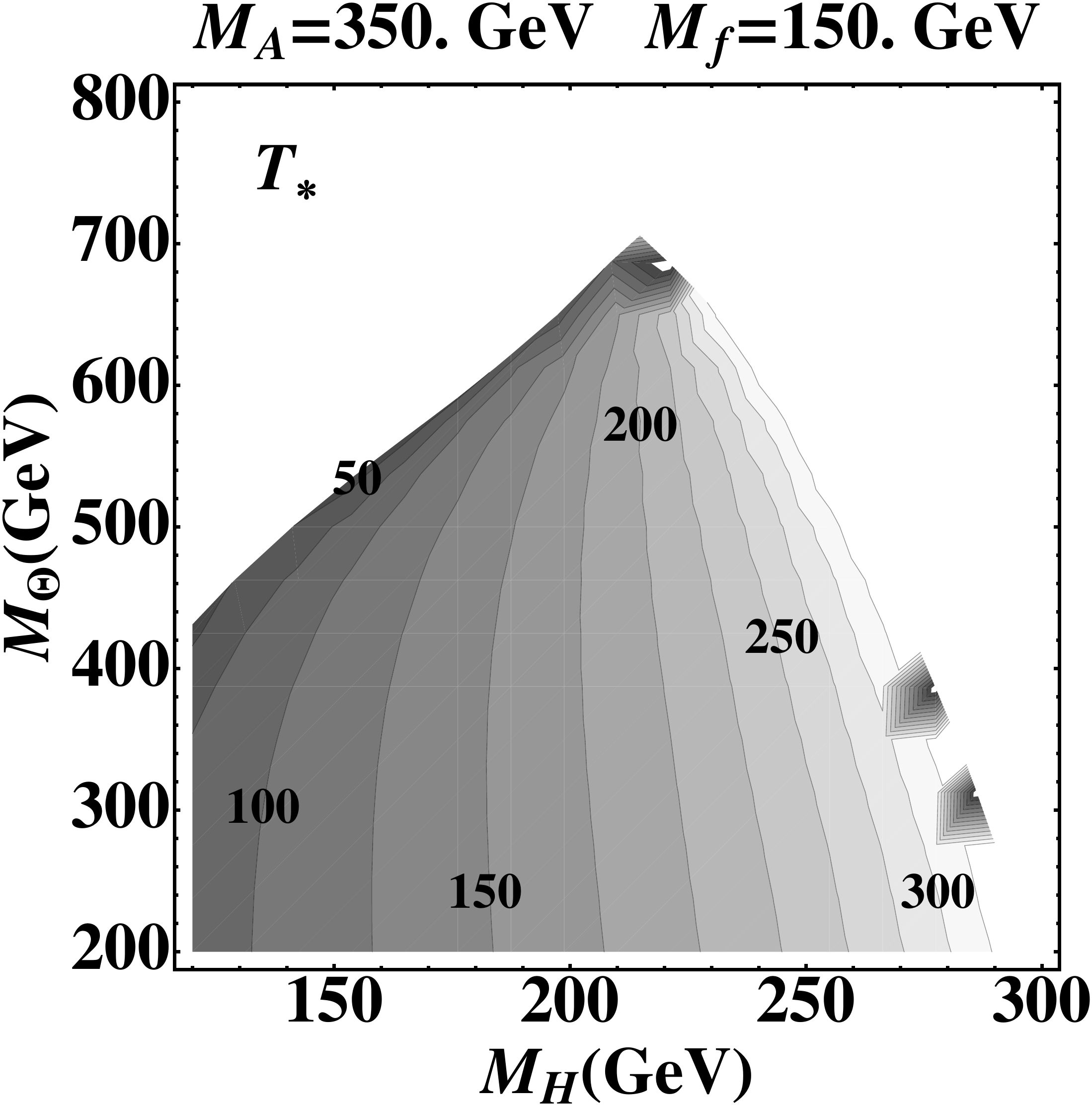}\hspace{0.1cm}\includegraphics[height=5.1cm,width=5.32cm]{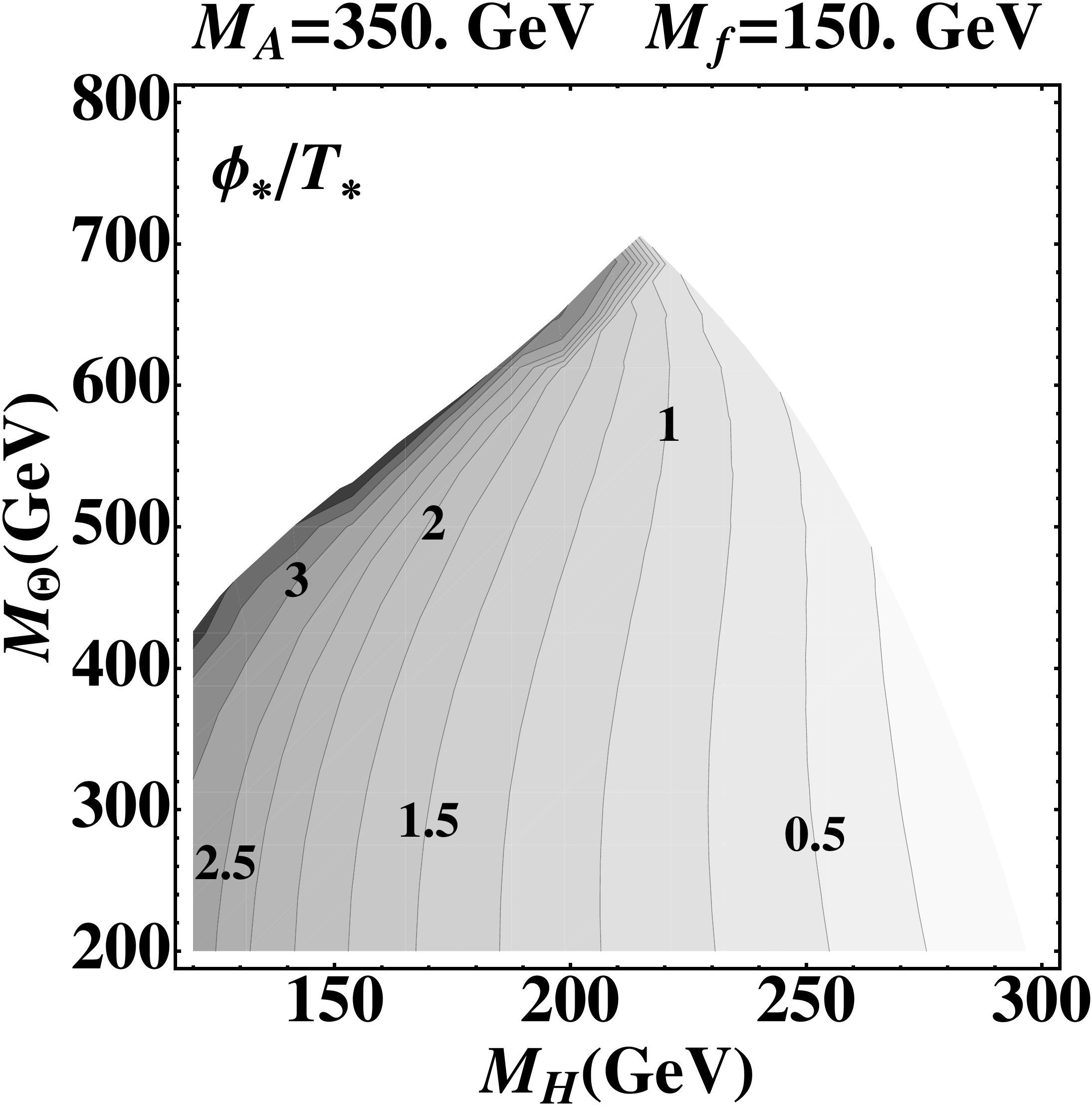}

\includegraphics[height=5.1cm,width=5.32cm]{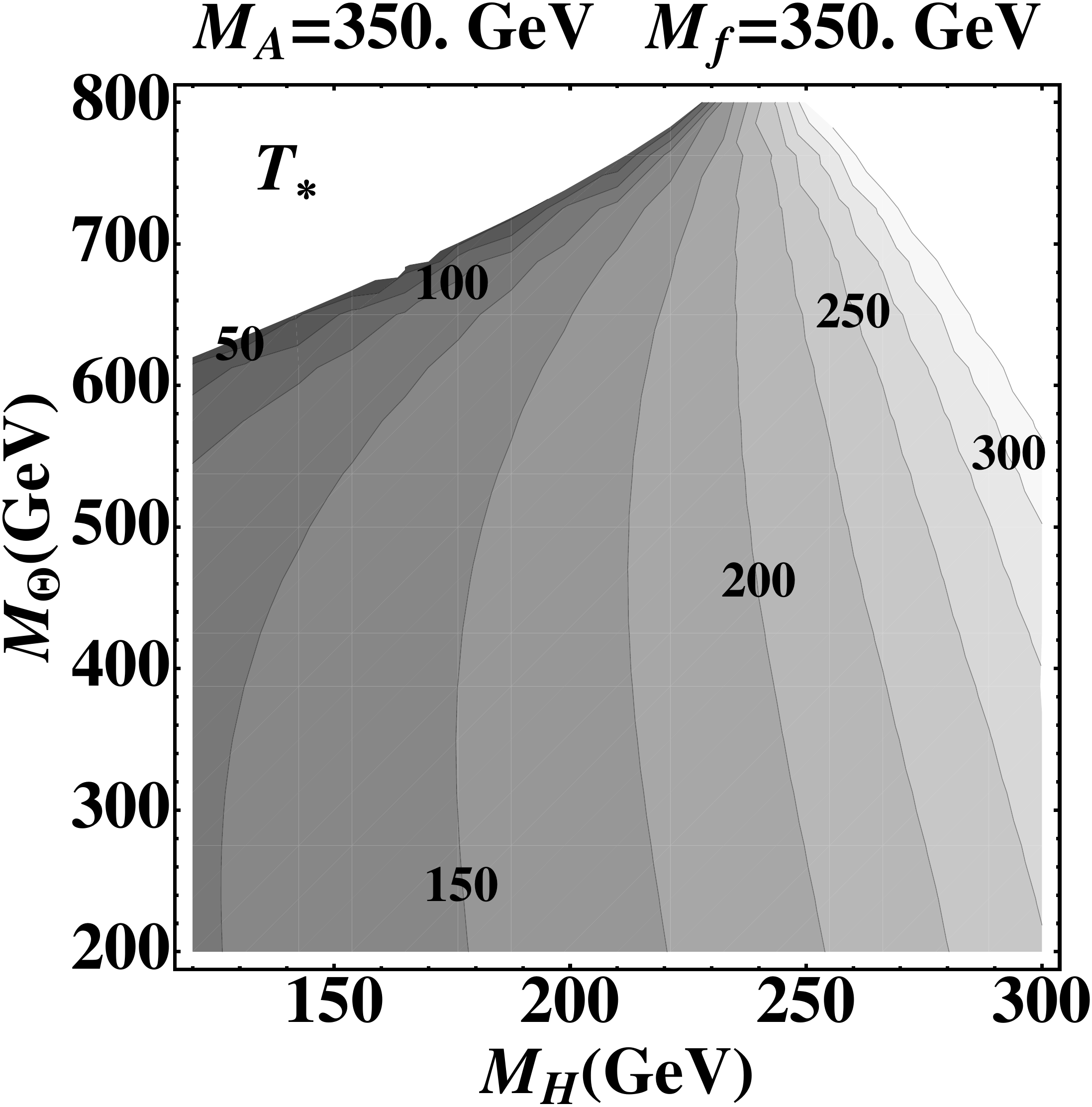}\hspace{0.1cm}\includegraphics[height=5.1cm,width=5.32cm]{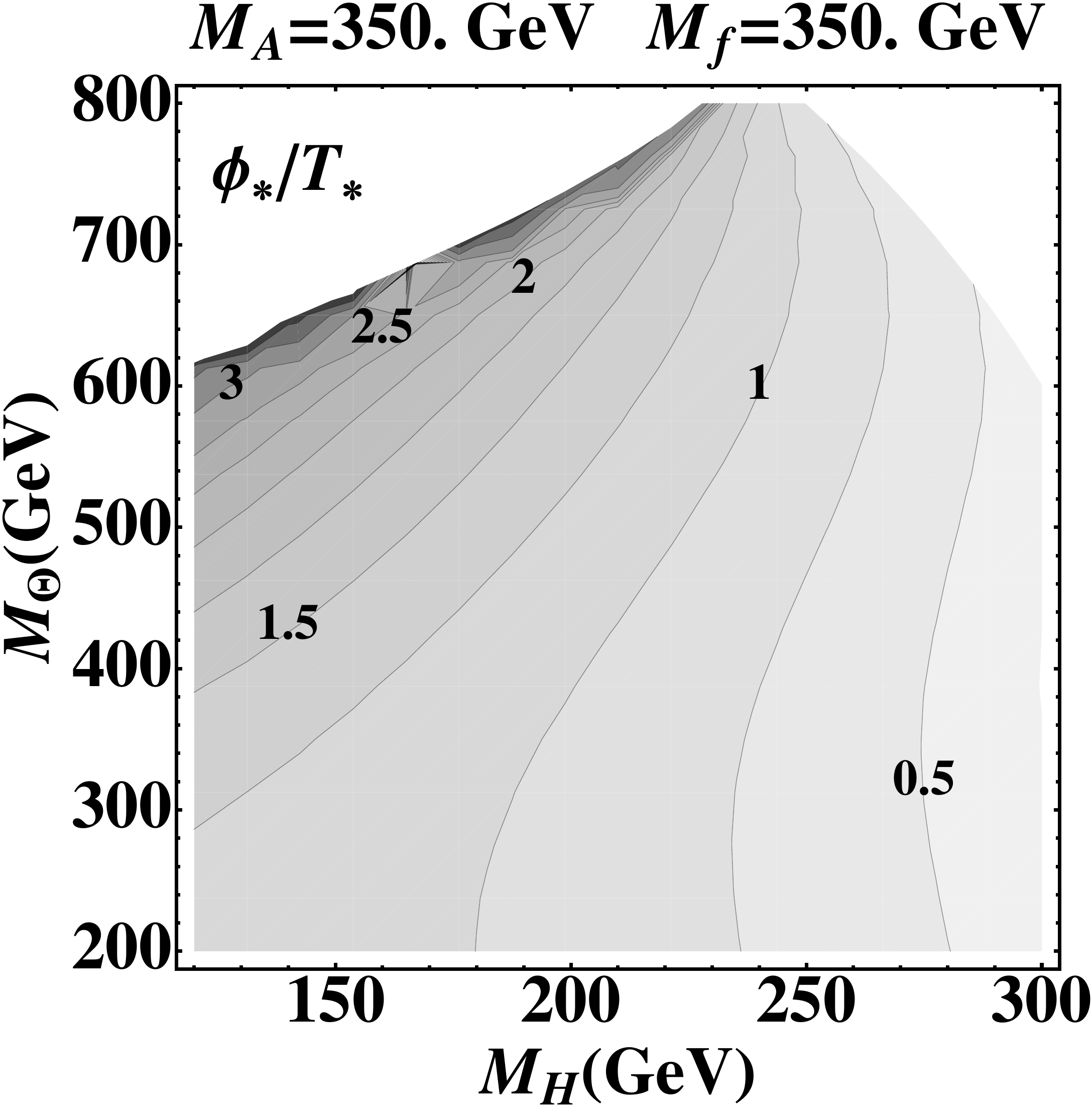}
}
\caption{The nucleation temperature $T_*$ (left column) and the strength of the transition $\phi_*/T_*$ (right column) for MWT in the
$M_H$-$M_{\Theta}$ plane for $M_A$, $M_{\rm f}=150$~GeV and $350$~GeV, as indicated in the labels. 
In the white regions the phase transition is either second order, very weakly first order, or does not occur at all.
}\label{fig:MWTresgen} \end{figure}

We include in the analysis the heaviest standard model particles, the top quark and the weak gauge bosons. In addition, we consider the fourth family leptons, and a few composite scalar states that are made of techniquarks. The scalar states are expected to be the lightest states of the technicolor theory and have masses near the electroweak scale with strong coupling to the chiral condensate, which is identified with the expectation value of the composite Higgs. Hence they are the most prominent states for the dynamics of the electroweak phase transition.
Of the two scenarios presented in \cite{Cline:2008hr} -- light and heavy ETC masses -- we only consider the latter one since it was seen to produce a stronger phase transition, potentially leading to stronger GWs. In this scenario the baryonic Goldstone bosons of the $SU(4) \to SO(4)$ chiral symmetry breaking 
are decoupled from the phase transition because of an ETC mass contribution that is much larger than the electroweak scale. The remaining eight scalar states include the composite Higgs $\sigma$ and its pseudoscalar partner $\Theta$ as well as the Goldstone bosons $\Pi$ that are eaten by the gauge bosons, and their scalar partners $A$.

The effective potential is then calculated as outlined in Appendix~\ref{app:effpot} and in \cite{Cline:2008hr}. We include the states listed above, except for the zero-temperature one-loop correction, where we leave out the (negligible but) infrared divergent contribution from the massless Goldstone bosons.

\subsection{Results}

\begin{figure}[ht]
 {\includegraphics[height=5.1cm,width=5.32cm]{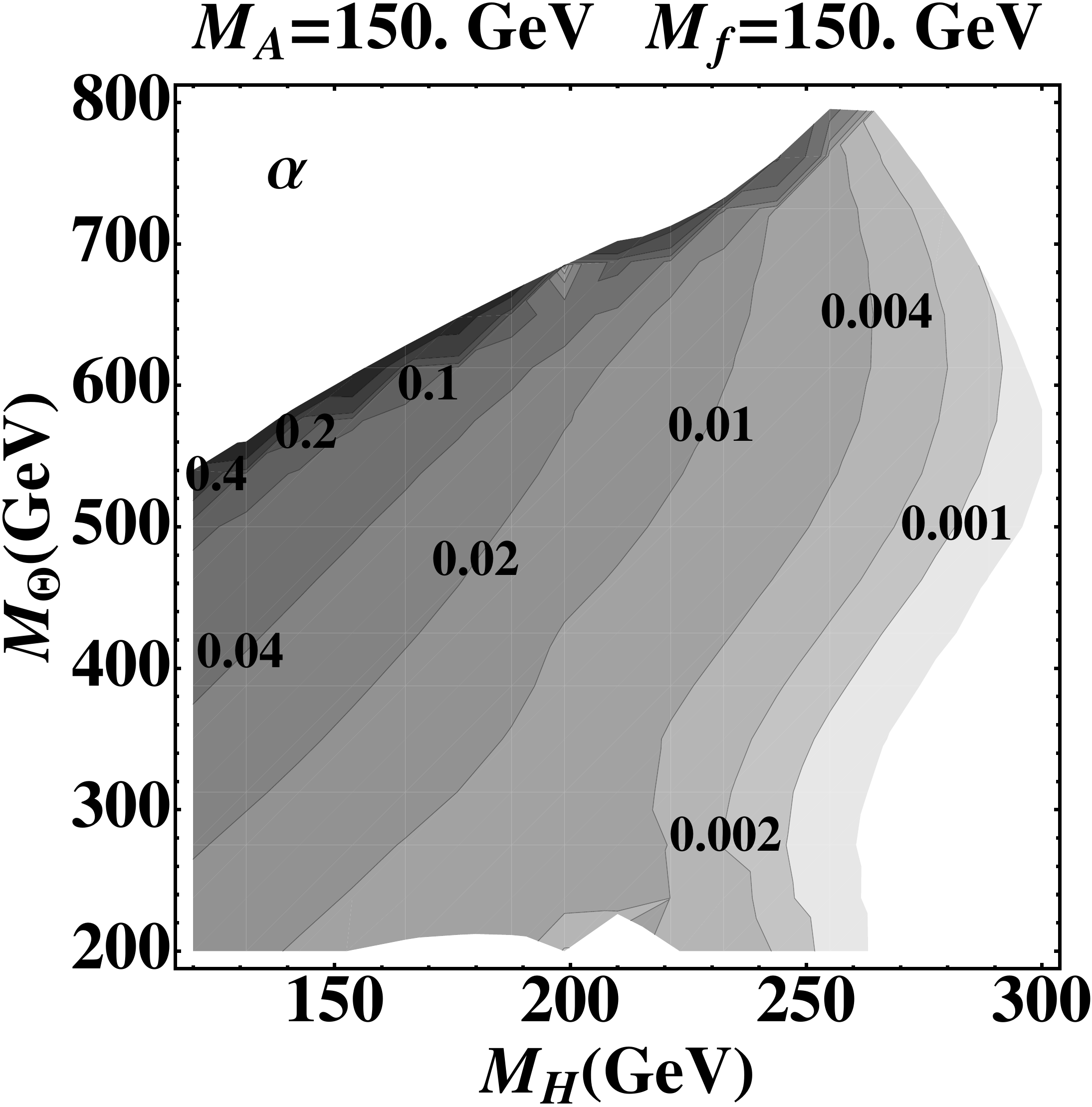}\hspace{0.1cm}\includegraphics[height=5.1cm,width=5.32cm]{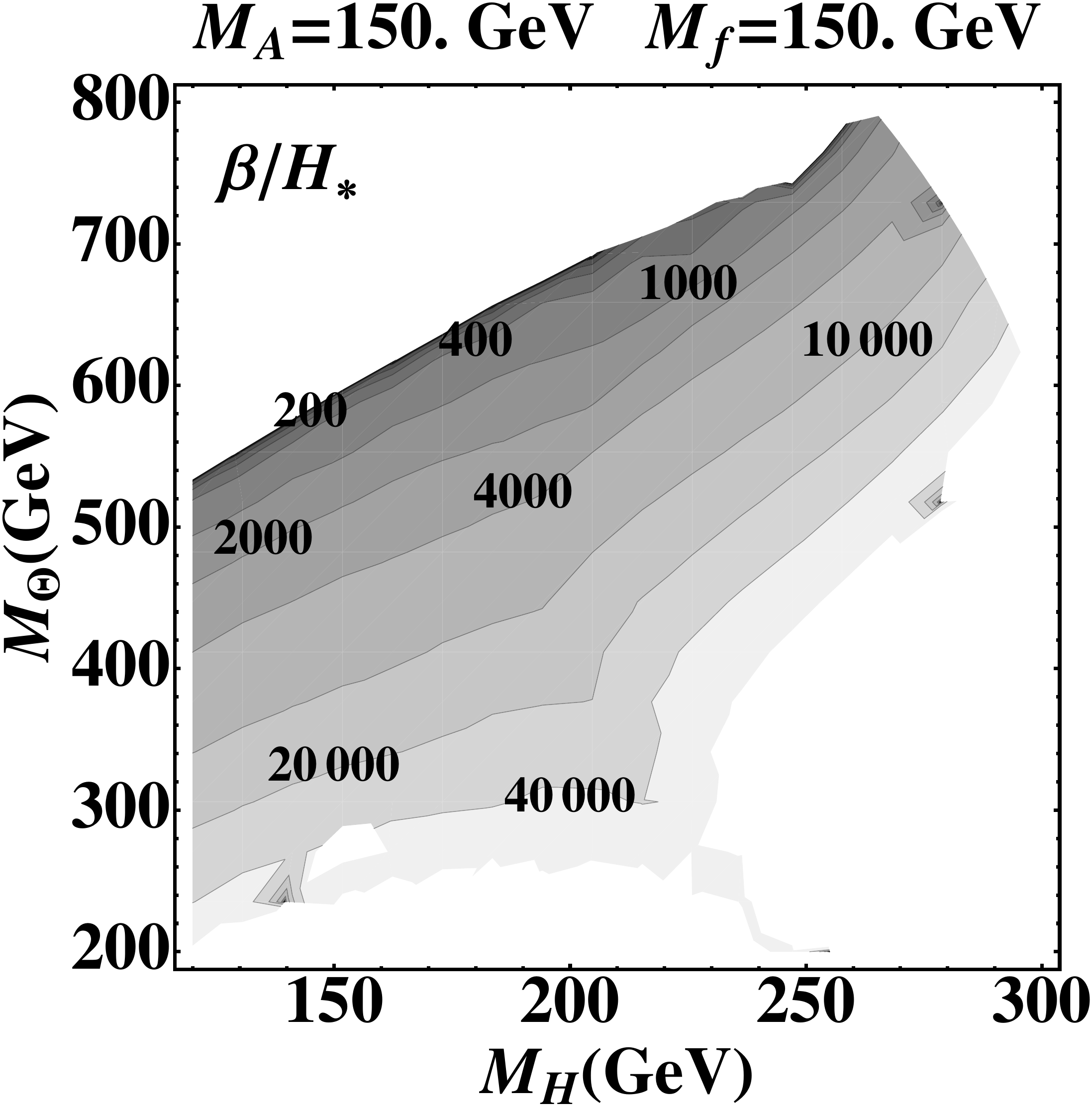}

\includegraphics[height=5.1cm,width=5.32cm]{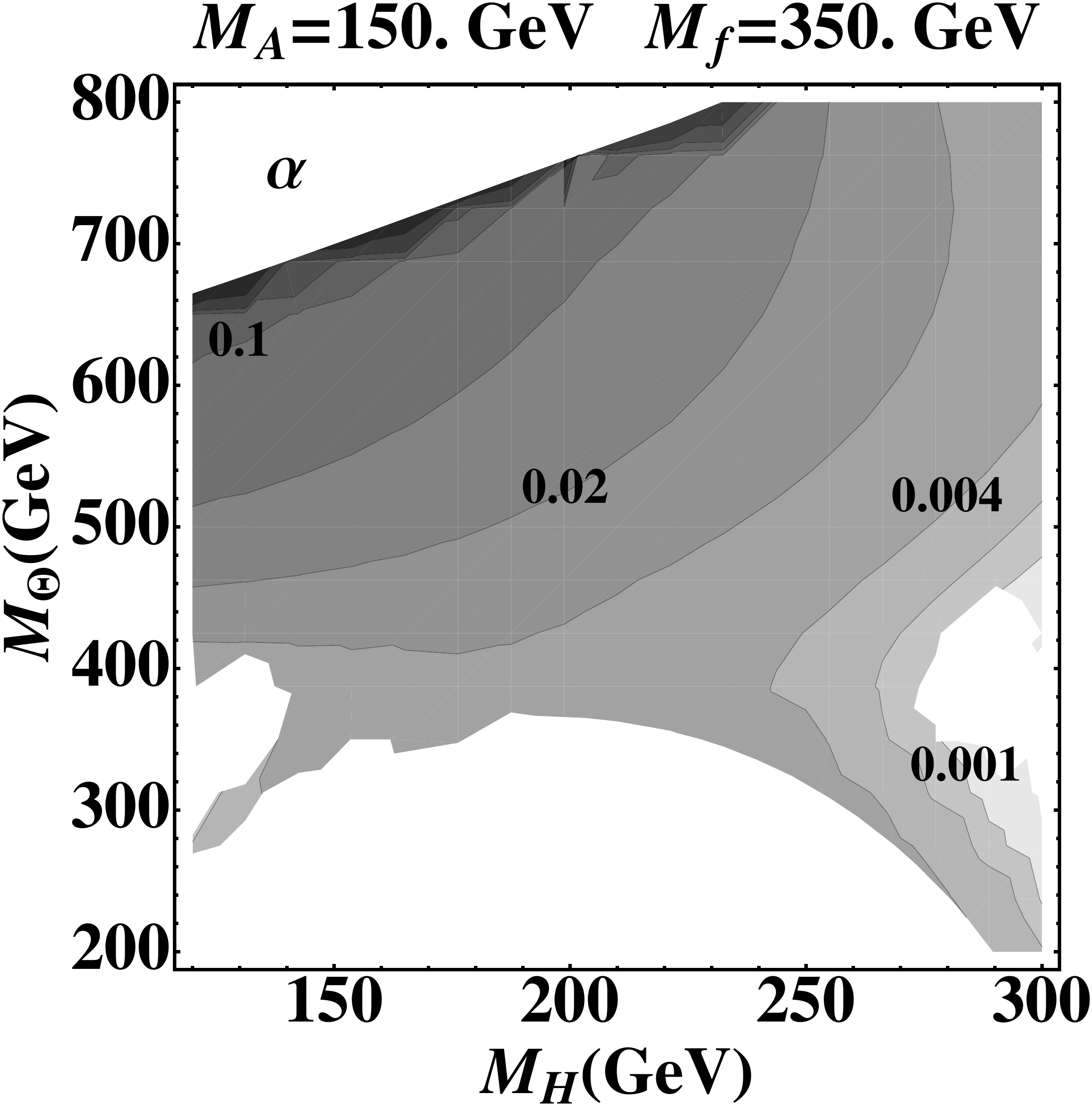}\hspace{0.1cm}\includegraphics[height=5.1cm,width=5.32cm]{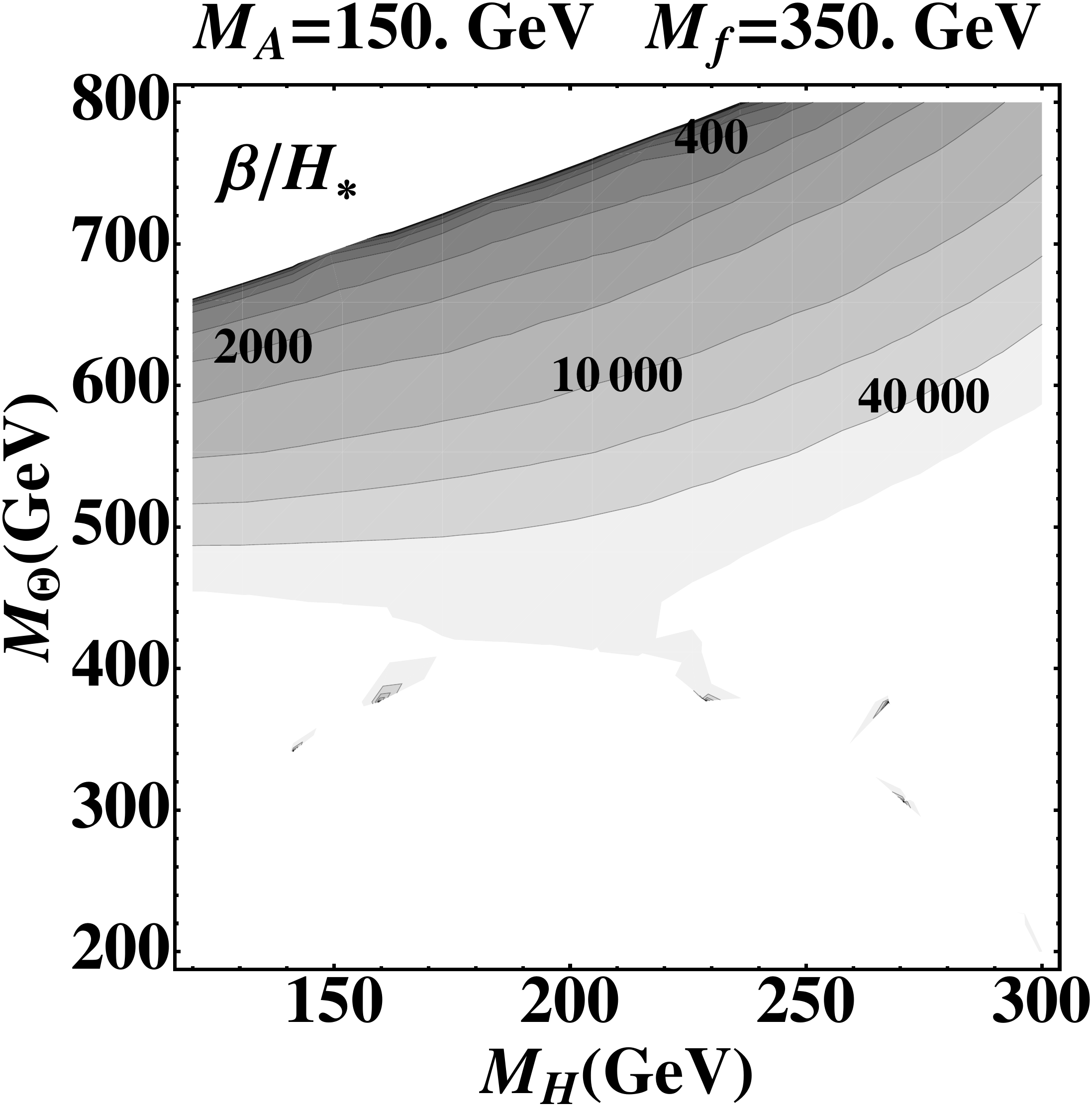}

\includegraphics[height=5.1cm,width=5.32cm]{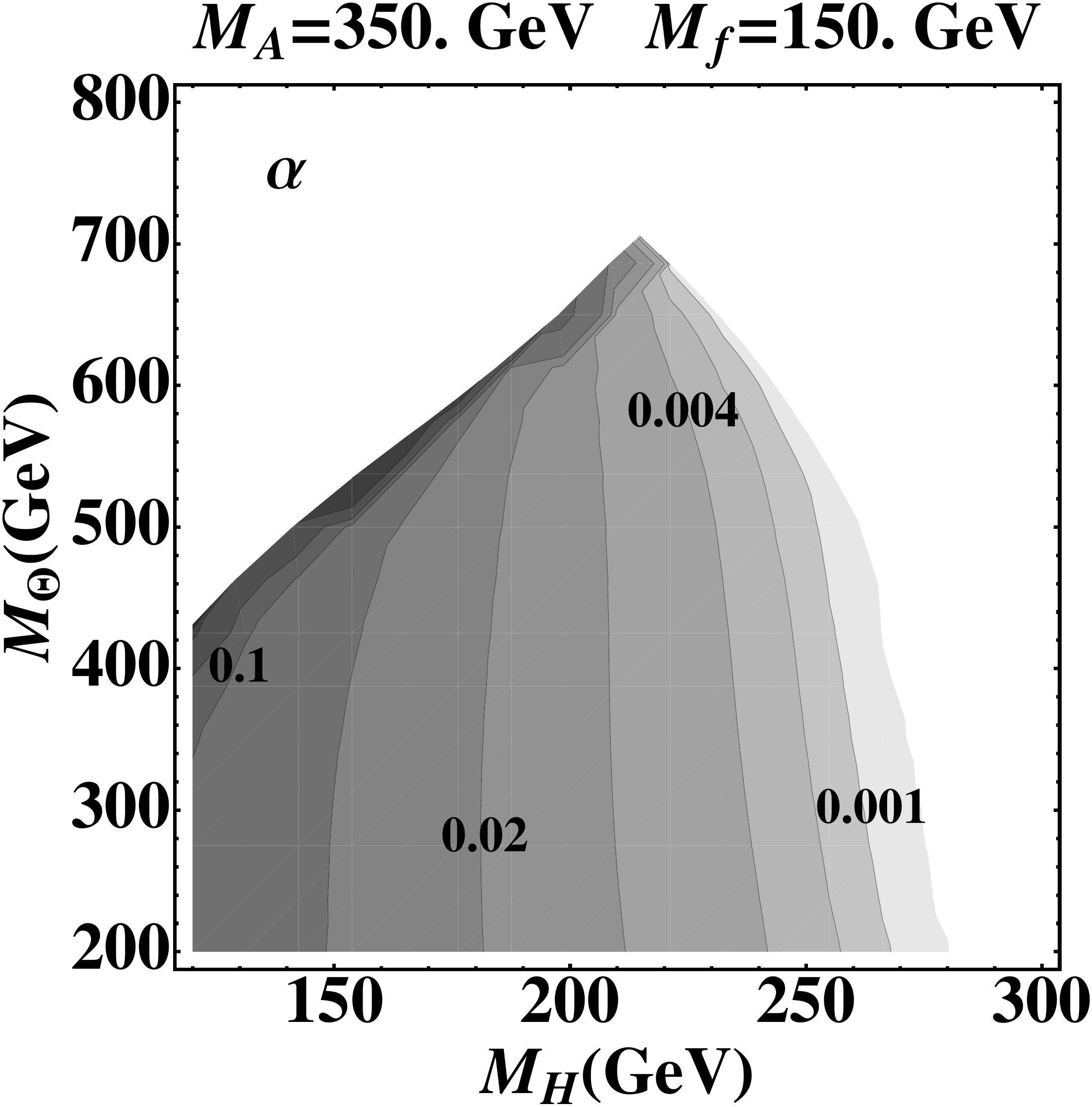}\hspace{0.1cm}\includegraphics[height=5.1cm,width=5.32cm]{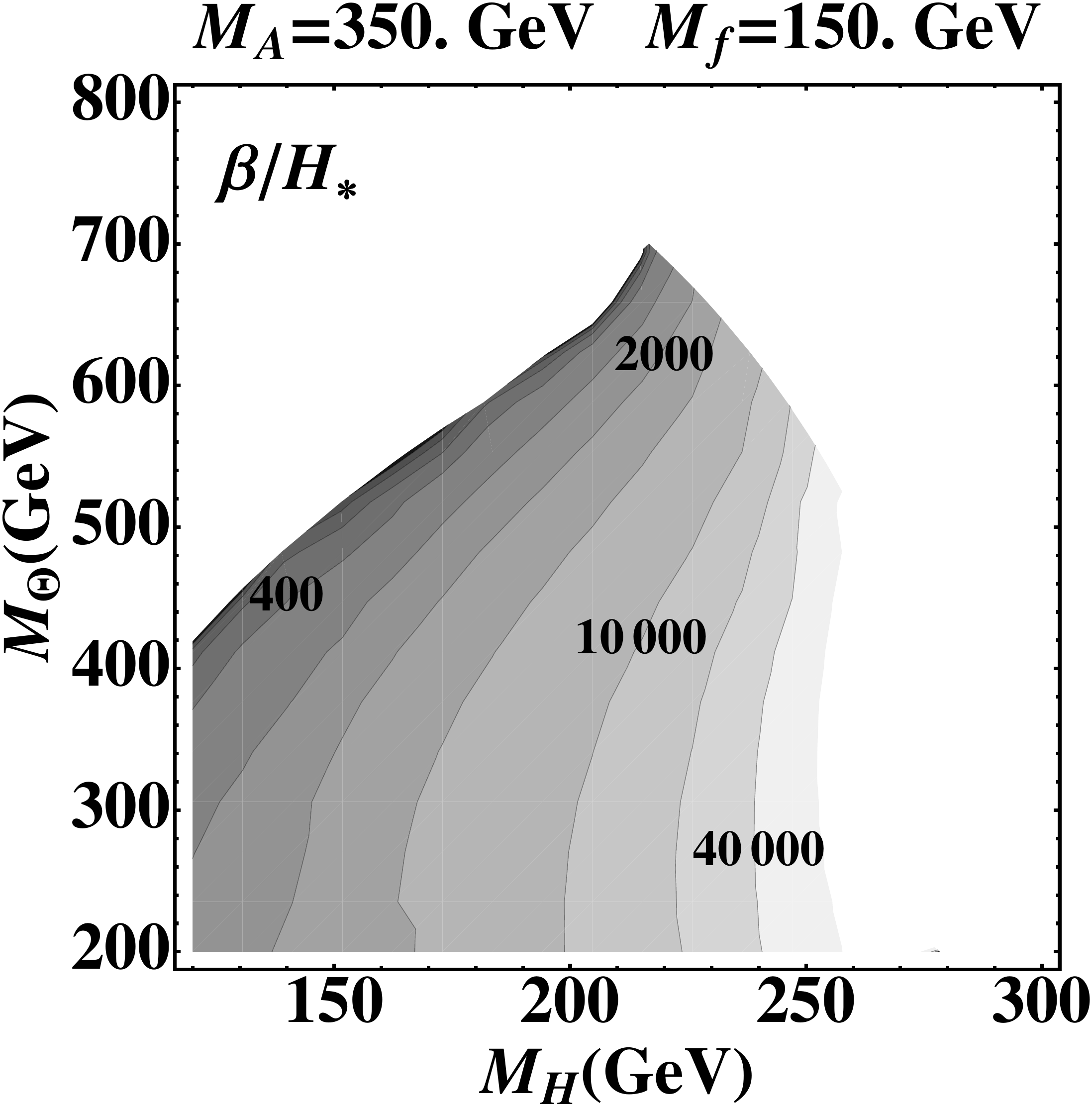}

\includegraphics[height=5.1cm,width=5.32cm]{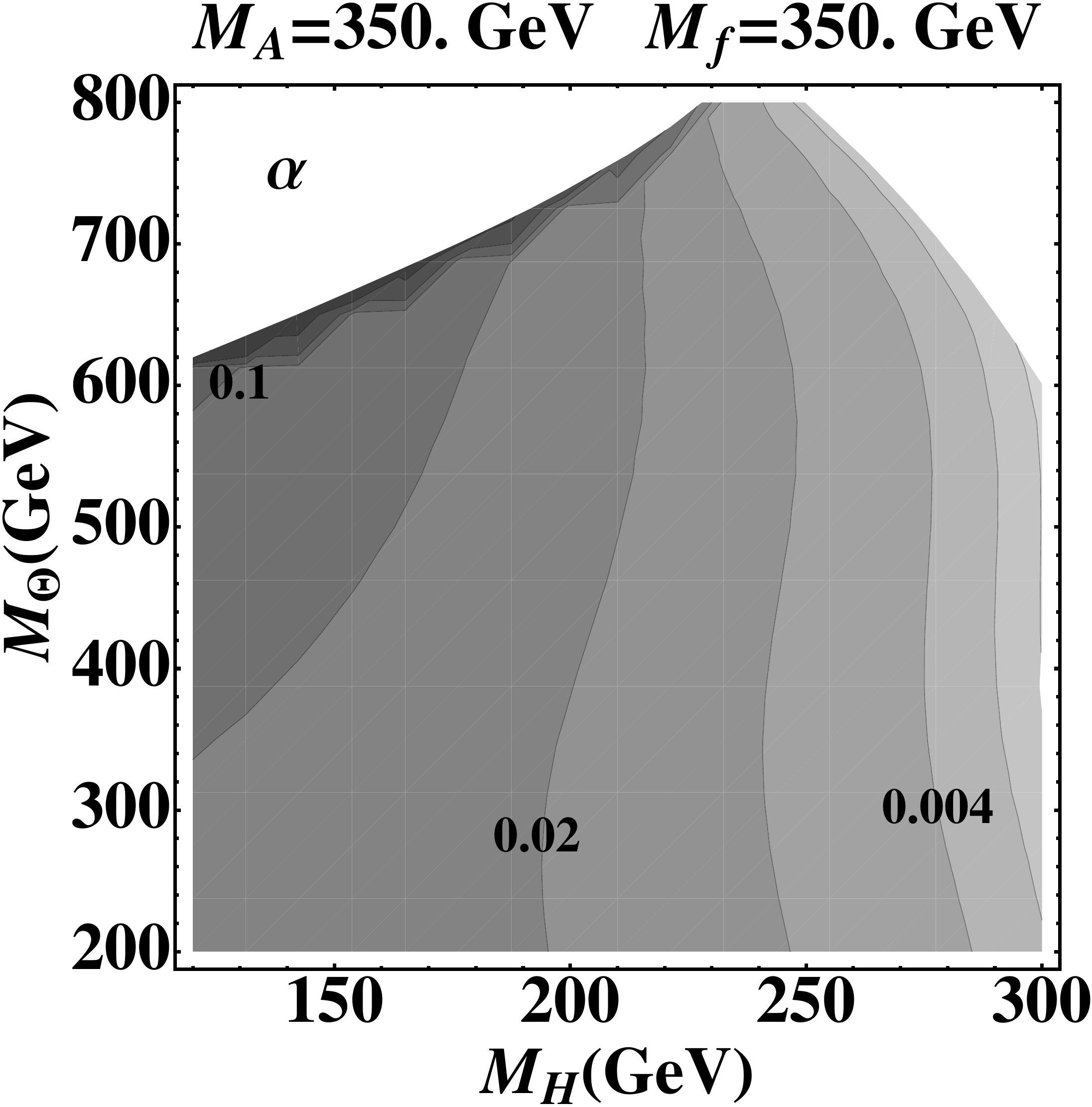}\hspace{0.1cm}\includegraphics[height=5.1cm,width=5.32cm]{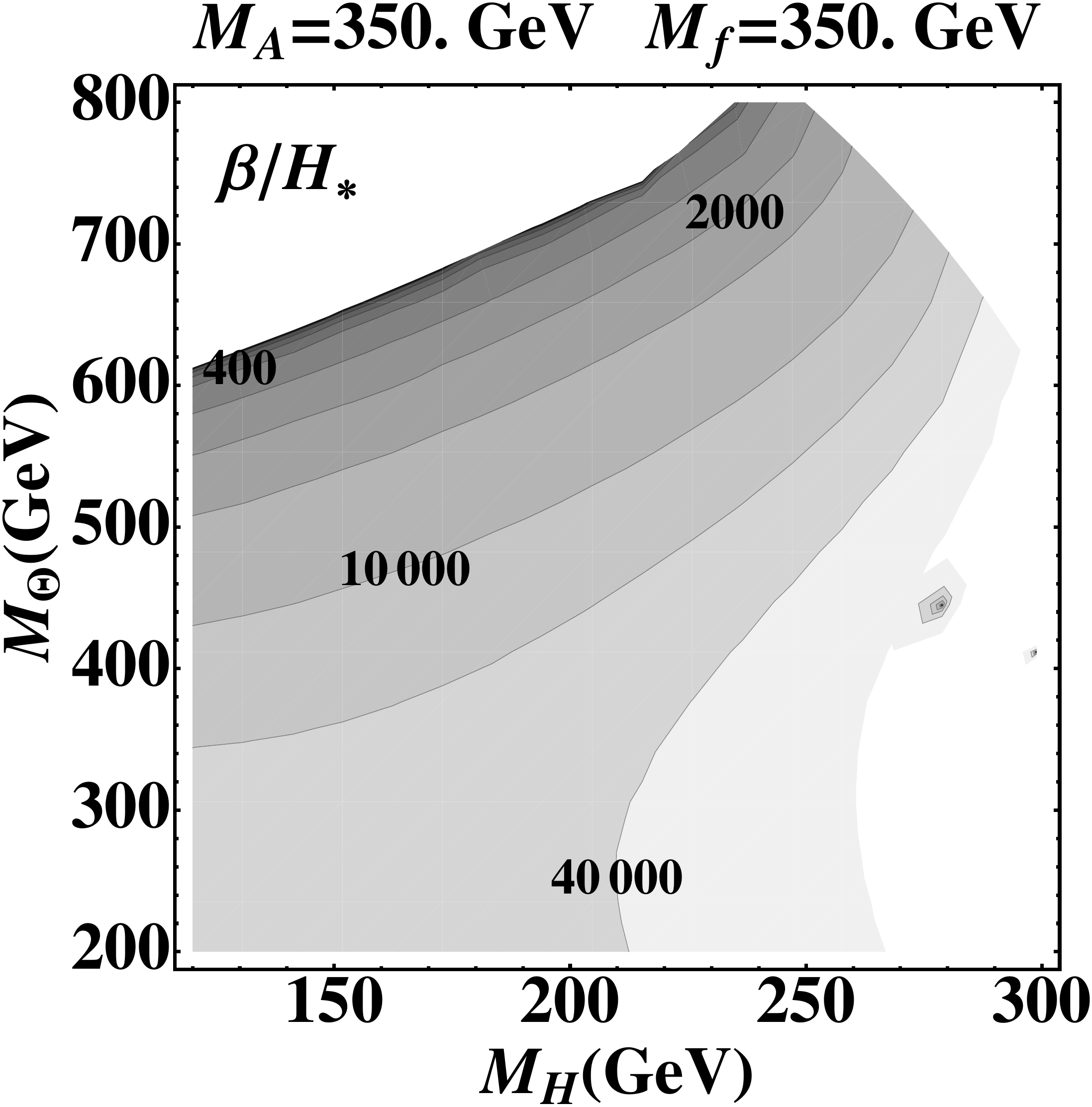}
}
\caption{The parameters $\alpha$ (left column) and $\beta/H_*$ (right column), which characterize the GW production, for MWT in the
$M_H$-$M_{\Theta}$ plane for $M_A$, $M_{\rm f}=150$~GeV and $350$~GeV, as indicated in the labels. 
}\label{fig:MWTres} \end{figure}

Let us first comment on the parameter space. One of the couplings $m^2$, $\lambda$, $\lambda'$, and $\lambda ''$ of the Higgs sector in the effective theory is fixed by the requirement $v=246$~GeV, while the rest can be expressed in terms of the Higgs mass $M_H$, the mass $M_\Theta$ of the $\Theta$ particle, and the mass $M_A$ of the scalar partners of the Goldstone bosons. Additional free parameters in our setup are the masses of the fourth family leptons which we assume to be equal and denote by $M_f$.

We follow \cite{Cline:2008hr} and present the results on the $M_H$ -- $M_\Theta$ plane while keeping $M_A$ and $M_f$ fixed at reference values $150$~GeV and $350$~GeV. Fig.~\ref{fig:MWTresgen} shows our results for the nucleation temperature $T_*$ (left) and the strength of the phase transition $\phi_*/T_*$.  We compared these results to our earlier estimate \cite{Cline:2008hr} for the strength $\phi_c/T_c$ of the phase transition at the critical temperature, where the two vacua are exactly degenerate. There is a notable difference only when the first order transition is very strong, $\phi_*/T_* \gtrsim 1$. In this case typically $\phi_*/T_* > \phi_c/T_c$. The region where the transition is strong enough to drive electroweak baryogenesis ($\phi/T \gtrsim 1$) is practically unchanged. 
Fig.~\ref{fig:MWTres} shows the results for the parameter $\alpha$ characterizing the produced latent heat (left), and the parameter $\beta/H_*$ characterizing the rapidness of the transition (right), which are specifically important for the production of GWs. 
It is seen that the values of the parameters are strongly correlated. As in other models (see, for example, \cite{Delaunay:2007wb}), strong first order transition, with sizeable $\phi_*/T_*$, generally means low $\beta/H_*$ and large $\alpha$, which are required for eminent production of GWs. The phase transition is at its strongest near a critical line on the $M_H$--$M_\Theta$ plane, where (within our approach) $T_* \to 0$ and $\phi_*/T_* \to \infty$. However, values of $\alpha\gtrsim 0.5$ (and $\beta/H_*\simeq 50 \ldots 1000$ ) that are required \cite{Grojean:2006bp} for the waves to be detectable at LISA \cite{Bender} are obtained only in a very narrow slice of the parameter space.

We observe that the value assumed by $\beta/H_*$  increases substantially as the strength of the phase transition decreases. We find, for strong phase transitions, agreement with the general prediction for this ratio put forward in \cite{Turner:1992tz,Kosowsky:1992vn} while for weak phase transitions we find agreement with the thin wall approximation.  
\begin{figure}[!tbp]
\begin{center}
\includegraphics[width=0.7\linewidth]{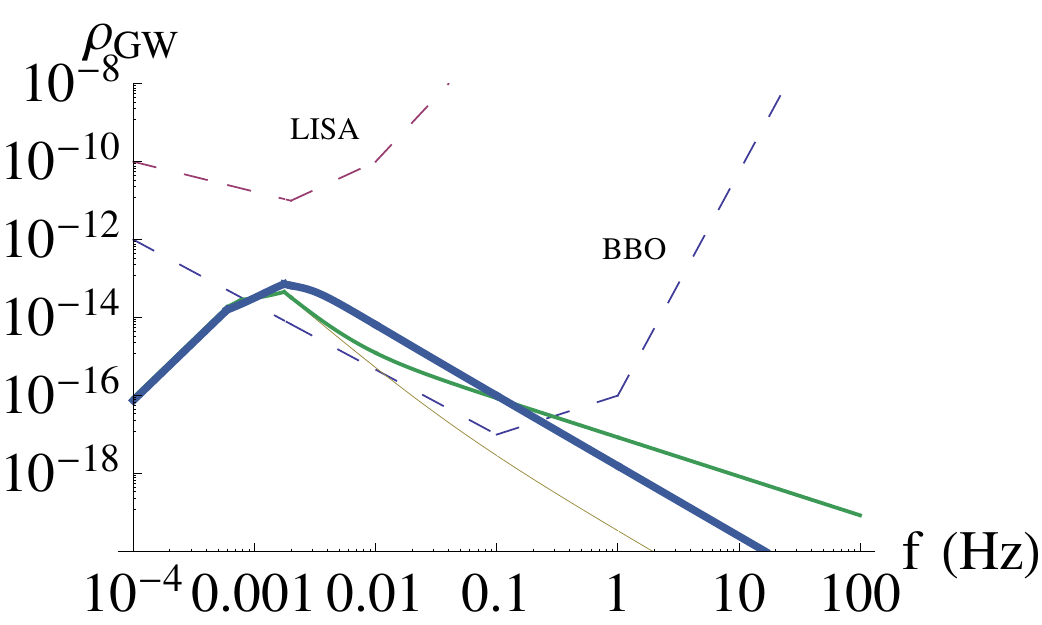}
\caption{The density of produced GWs $\rho_{\text{GW}}=\Omega_{\text{GW}} \text{h}^2$ in the case of the MWT as a function of the frequency (in Hz). Dashed lines represent the expected sensitivity of LISA and BBO, while solid lines represent the gravitational spectrum of bubble collision and turbulence combined. The three solid lines from thinner to thicker correspond to the gravitational spectrum with bubble collisions given, respectively, by Eqs.~(\ref{col1}),~(\ref{col2}),~(\ref{col3}). The values of the parameters are given in the text.} \label{fig:gw_mwt}
\end{center}
\end{figure}

In Fig.~\ref{fig:gw_mwt} we have plotted the gravitational spectrum of a MWT theory with $\alpha=0.2$, $\beta/H_*=300$ and $T_*=60$ GeV. This is an example of a strong first order phase transition with $\phi_*/T_* \simeq 4$, which can be obtained without fine-tuning the parameter values to be unnaturally close to the critical line. This set of numbers can be derived from the effective potential if we choose, for example, $M_A=M_f=150$ GeV, $M_H\simeq150$ GeV, and $M_\Theta\simeq 583$ GeV. For these values of $M_H$, $M_A$, and $M_f$ the phase transition disappears for $M_\Theta \gtrsim 592$ GeV, so the chosen value of $M_\Theta$ lies within $10$~GeV from the critical line.

\section{Gravitational waves from multiple phase transitions}

An interesting scenario which may arise in strongly interacting extensions of the standard model is that there can be several phase transitions at temperatures close to the electroweak scale \cite{Jarvinen:2009wr}. Only one of these transitions needs to break the electroweak symmetry, while all of them can produce GWs, which can lead to a complex GW spectrum with several peaks from the various transitions. We shall consider here the case of Ultra Minimal Technicolor (UMT) \cite{Ryttov:2008xe}, where chiral symmetry breaking can proceed in two (or possibly three \cite{Jarvinen:2009wr}) steps, related to the two sectors of matter in this theory. 

In general, the two sectors of UMT can talk to each other, since the matter in the different sectors interact via the strong technicolor dynamics. This interplay is realized at the effective Lagrangian level by terms which involve composite bound states from both sectors. To simplify the discussion on the production of GWs, we shall here omit these terms (set $\delta=0=\delta'$ below) and assume that the two sectors are decoupled. Then the electroweak gauge symmetry feels only one of the sectors (the one with the order parameter $\sigma_4$ below), while the other (the one with the order parameter $\sigma_2$) is decoupled from the electroweak dynamics.
In this case, the formalism for GW production presented above is directly applicable for UMT. 
The effect of the interactions between the two sectors was studied in \cite{Jarvinen:2009wr,Jarvinen:2009pk}. The interactions were found to lead to a rich phase diagram with the possibility of breaking the electroweak symmetry twice (and restoring it once) as the universe cools down, while the phase transitions were typically weaker than in the decoupled case. Hence we expect that also the produced GWs are at their strongest in the scenario investigated here. However, generally even within UMT, there is the possibility of having three phase transitions and consequently a richer GW spectrum than the one presented here.

Let us make a brief general comment on multiple (first order) transitions in the early universe. An important question is if the transitions can be treated separately or if there are bubbles related to different transitions present simultaneously. Recall from the discussion above that the nucleation probability density is $\sim T^4e^{-S_E}$ and consequently the time scale of nucleation is given by the inverse of $\beta \equiv - dS_E/dt$. The scale of change in temperature is thus
\be \Delta T \sim  \frac{1}{\beta}\frac{dT}{dt} \sim T_* \frac{H_*}{\beta} \ . \ee
Since we will have $\beta/H_* > 100$ (typically even $\beta/H_* > 1000$) and $T_*$ is around a few hundred GeV, $\Delta T$ is less than a few GeV. Hence the transitions can be practically treated as separate, unless the underlying dynamics forces them to be exactly simultaneous \cite{Jarvinen:2009wr,Jarvinen:2009pk}. Notice that the nucleation probability depends exponentially on $S_E$; therefore the transition will end very quickly when the temperature difference with respect to the start of nucleation exceeds $T_* H_*/\beta$.

\subsection{Effective theory for UMT}

The model proposed in \cite{Ryttov:2008xe} consists of an $SU(2)$ gauge group with two Dirac fermions belonging to the fundamental representation and two Weyl fermions belonging to the adjoint representation. In order not to be in conflict with the Electroweak Precision Tests only the fundamental fermions are charged under the electroweak symmetry.


We shall only consider the effect of the composite scalar mesons which are expected to be the lightest particles in the theory. Their masses have the strongest dependence on the vacuum expectations values of the Higgs fields.

The relevant degrees of freedom are efficiently collected in two distinct matrices, $M_4$ and $M_2$, which transform as $M_4 \rightarrow g_4M_4g_4^T$ and $M_2 \rightarrow g_2M_2g_2^T$ with $g_4 \in SU(4)$ and $g_2 \in SU(2)$. Both $M_4$ and $M_2$ consist of a composite isoscalar and its pseudoscalar partner together with the Goldstone bosons and their scalar partners
\begin{eqnarray}
M_4 &=& \left[ \frac{\sigma_4 + i \Theta_4}{2} + \sqrt{2}\left( i \Pi_4^i+ \tilde{\Pi}_4^i \right) X_4^i \right] E_4 \ , \qquad i=1,\ldots,5 \ , \\
M_2 &=& \left[ \frac{\sigma_2 + i \Theta_2}{\sqrt{2}} + \sqrt{2} \left( i \Pi_2^i+ \tilde{\Pi}_2^i \right) X_2^i \right] E_2 \ , \qquad i=1,2 \ .
\end{eqnarray}

The notation is such that $X_{4}$ and $X_{2}$ are the broken generators of $SU(4)$ and $SU(2)$ respectively. 
Also $\sigma_{4}$ and $\Theta_{4}$ are the composite Higgs and its pseudoscalar partner while $\Pi_{4}^i$ and $\tilde{\Pi}_{4}^i$ are the Goldstone bosons and their associated scalar partners. For $SU(2)$ one simply substitutes the index $4$ with the index $2$.

To describe the interaction with the weak gauge bosons we embed the electroweak gauge group in $SU(4)$ as done in \cite{Appelquist:1999dq}.
Because of the choice of the electroweak embedding the weak interactions explicitly reduce the $SU(4)$ symmetry to $SU(2)_L \times U(1)_Y \times U(1)_{TB}$ which is further broken to $U(1)_{\rm em} \times U(1)_{TB}$ via the technicolor interactions. $U(1)_{TB}$ is the technibaryon number related to the fundamental fermions. 
The remaining $SU(2)\times U(1)$ spontaneously breaks, via the extra technifermion condensates, to $SO(2)\times Z_2$. Here $SO(2)\cong U(1)$ is the technibaryon number related to the adjoint fermions.

We are now in a position to write down the effective Lagrangian. It contains the kinetic terms and a potential term
\begin{eqnarray} \label{Leff}
\mathcal{L} &=& \frac{1}{2} \text{Tr} \left[ D_{\mu} M_4 D^{\mu} M_4^{\dagger} \right] + \frac{1}{2} \text{Tr} \left[ \partial_{\mu} M_2 \partial^{\mu} M_2^{\dagger} \right] - \mathcal{V} \left( M_4, M_2 \right)  \ ,
\end{eqnarray}
where the potential is
\begin{eqnarray}
\mathcal{V} \left( M_4, M_2 \right) &=& -\frac{m_4^2}{2} \text{Tr}\left[ M_4M_4^{\dagger} \right] + \frac{\lambda_4}{4}\text{Tr}\left[ M_4M_4^{\dagger} \right]^2 + \lambda_4' \text{Tr} \left[ M_4M_4^{\dagger}M_4M_4^{\dagger} \right] \nonumber \\
&&
-\frac{m_2^2}{2} \text{Tr}\left[ M_2M_2^{\dagger} \right] + \frac{\lambda_2}{4}\text{Tr}\left[ M_2M_2^{\dagger} \right]^2 + \lambda_2' \text{Tr} \left[ M_2M_2^{\dagger}M_2M_2^{\dagger} \right] \\
&&
+ \frac{\delta}{2}\text{Tr}\left[ M_4M_4^{\dagger} \right] \text{Tr}\left[ M_2M_2^{\dagger} \right] +4\delta' \left[ \left( \det M_2 \right)^2 \text{Pf}\ M_4 + \text{h.c.} \right] \ . \nonumber
\end{eqnarray}
We shall from now on set $\delta=0=\delta'$ so that the two sectors are decoupled.

Once $M_4$ develops a vacuum expectation value the electroweak symmetry breaks and three of the eight Goldstone bosons - $\Pi^0,\ \Pi^+$, and $\Pi^-$ - will be eaten by the massive gauge bosons. In terms of the parameters of the theory the vacuum states $\langle \sigma_4 \rangle = v_4$ and $\langle \sigma_2 \rangle = v_2$ which minimize the potential are 
\begin{eqnarray}\label{vacua1}
m_4^2 &=&  \left( \lambda_4 + \lambda_4' \right) v_4^2 \ , \\
m_2^2 &=&  \left( \lambda_2 + 2\lambda_2' \right) v_2^2 \ . \label{vacua2}
\end{eqnarray}

For the model to be phenomenologically viable some of the Goldstone bosons must acquire a mass. Here we parametrize the ETC interactions by adding at the effective Lagrangian level the operators needed to give the unwanted Goldstone bosons an explicit mass term.

The effective ETC Lagrangian breaks the global $SU(4)\times SU(2) \times U(1)$ symmetry.
To construct the required ETC terms at the effective Lagrangian level, we find it useful to split
$M_4$ ($M_2$) -- form invariant under $U(4)$ ($U(2)$) --  as follows:
\begin{equation}
M_4 = \tilde{M}_4  + i P_4 \ , \qquad  {\rm and} \qquad M_2 = \tilde{M}_2  + i P_2 \ ,
\end{equation}
with
\begin{eqnarray}
\tilde{M}_4 &=& \left[ \frac{\sigma_4}{2} + i\sqrt{2}  \Pi_4^i   X_4^i \right]   E_4 \ , \quad P_4 = \left[ \frac{ \Theta_4}{2}  -i\sqrt{2} \tilde{\Pi}_4^i X_4^i \right] E_4 \ , \qquad i=1,\ldots,5 \ , \\
\tilde{M}_2 &=& \left[ \frac{\sigma_2 }{\sqrt{2}} + i\sqrt{2}  \Pi_2^i  X_2^i \right] E_2 \ , \quad P_2 = \left[ \frac{ \Theta_2}{\sqrt{2}}  -i\sqrt{2} \tilde{\Pi}_2^i  X_2^i \right] E_2 \ , \qquad i=1,2 \ .\end{eqnarray}
$\tilde{M}_4 $ ($\tilde{M}_2$) as well as $P_4$ ($P_2$) are separately $SU(4)$ ($SU(2)$) form invariant. A set of operators able to give masses to the electroweak neutral Goldstone bosons is
\begin{eqnarray} \label{LETC}
\mathcal{L}_{ETC} &=&  \frac{m_{4,ETC}^2}{4} \text{Tr} \left[ \tilde{M}_4B_4 \tilde{M}_4^{\dagger} B_4 + \tilde{M}_4 \tilde{M}_4^{\dagger} \right] + \frac{m_{2,ETC}^2}{4} \text{Tr} \left[ \tilde{M}_2B_2 \tilde{M}_2^{\dagger} B_2 + \tilde{M}_2 \tilde{M}_2^{\dagger} \right] \nonumber \\
&& - m_{1,ETC}^2 \left[ \text{Pf}\ P_4 + \text{Pf}\ P_4^{\dagger} \right] - \frac{m_{1,ETC}^2}{2} \left[ \det(P_2) + \det(P_2^{\dagger}) \right] \ ,
\end{eqnarray}
where $B_4$ is the diagonal $SU(4)$ generator that commutes with the electroweak generators, and $B_2 = \tau^3$ is the diagonal generator of $SU(2)$. The masses of the two Higgs particles
\begin{eqnarray}
  M^2_{H_4} &=& 2 m_4^2 = 2 \left( \lambda_4 + \lambda_4' \right) v_4^2 \\
  M^2_{H_2} &=& 2m_2^2  = 2 \left( \lambda_2 + 2\lambda_2' \right) v_2^2
\end{eqnarray}
are unaffected by the addition of the ETC low energy operators.
The rest of the spectrum is:
\begin{eqnarray}
M^2_{\Pi_{UD}} = m_{4,ETC}^2 \ , \qquad M^2_{\Pi_{\lambda\lambda}} = m_{2,ETC}^2 \ , \qquad M^2_{\Theta_4} = m_{1,ETC}^2 = M_{\Theta_2}^2
\end{eqnarray}
for the pseudoscalar partners and the Goldstone bosons that are not eaten by the massive vector bosons and:
\begin{eqnarray}
M^2_{\tilde{\Pi}_{UD}} &=& M^2_{\tilde{\Pi}^0} = M^2_{\tilde{\Pi}^{\pm}} = 2\left( \lambda'_4 v_4^2 + \delta' v_2^4 \right) + m^2_{1,ETC} \ , \\
M^2_{\tilde{\Pi}_{\lambda\lambda}} &=& 4v_2^2 \left( \lambda_2' + \delta' v_4^2 \right) + m^2_{1,ETC} \ , 
\end{eqnarray}
for the 
scalar partners.

\subsection{Effective potential analysis for UMT} 

For the effective potential in UMT we use the same approach as outlined above for MWT: we include one-loop corrections with ring resummation for the bosonic degrees of freedom as presented in Appendix~\ref{app:effpot} and in \cite{Jarvinen:2009pk}. The particle spectrum includes the top quark, the weak gauge bosons, and the lowest scalar states of the theory presented above. For UMT the ETC mass scale must be of the order of the electroweak scale in order to obtain first order transitions.

\begin{figure}[ht]
 {\includegraphics[height=5.11cm,width=5.32cm]{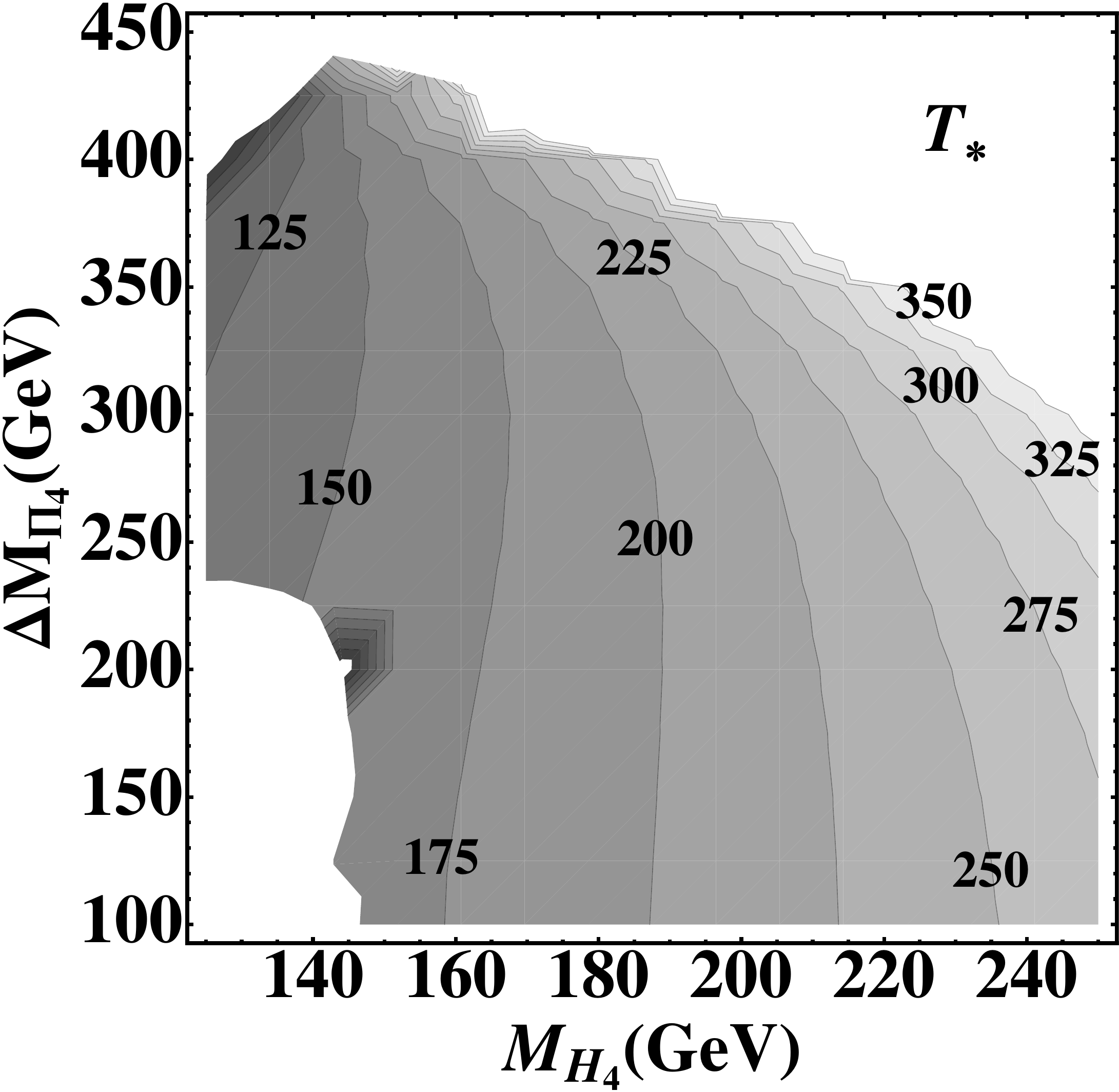}\hspace{0.1cm}\includegraphics[height=5.11cm,width=5.32cm]{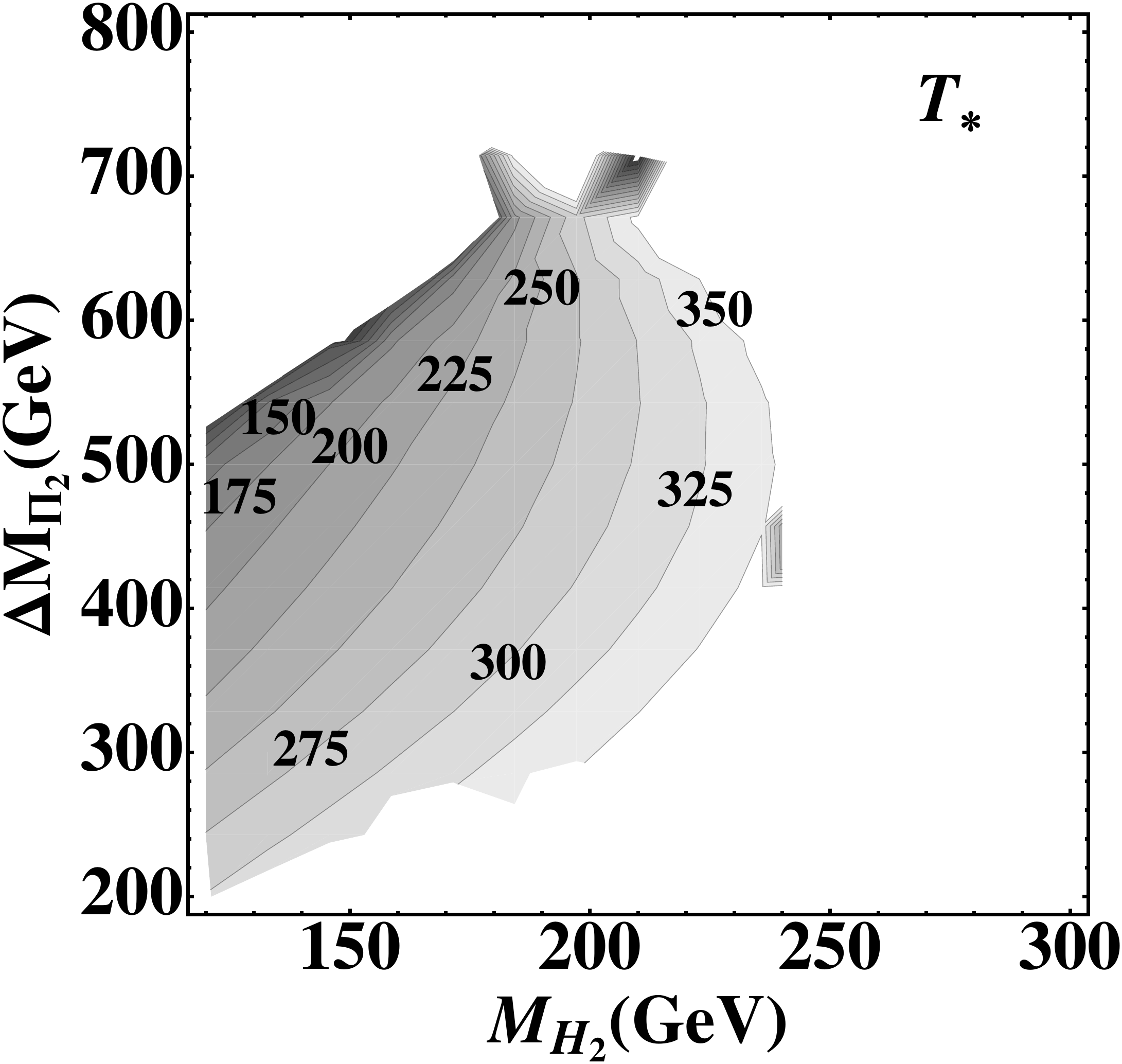}

\includegraphics[height=5.11cm,width=5.32cm]{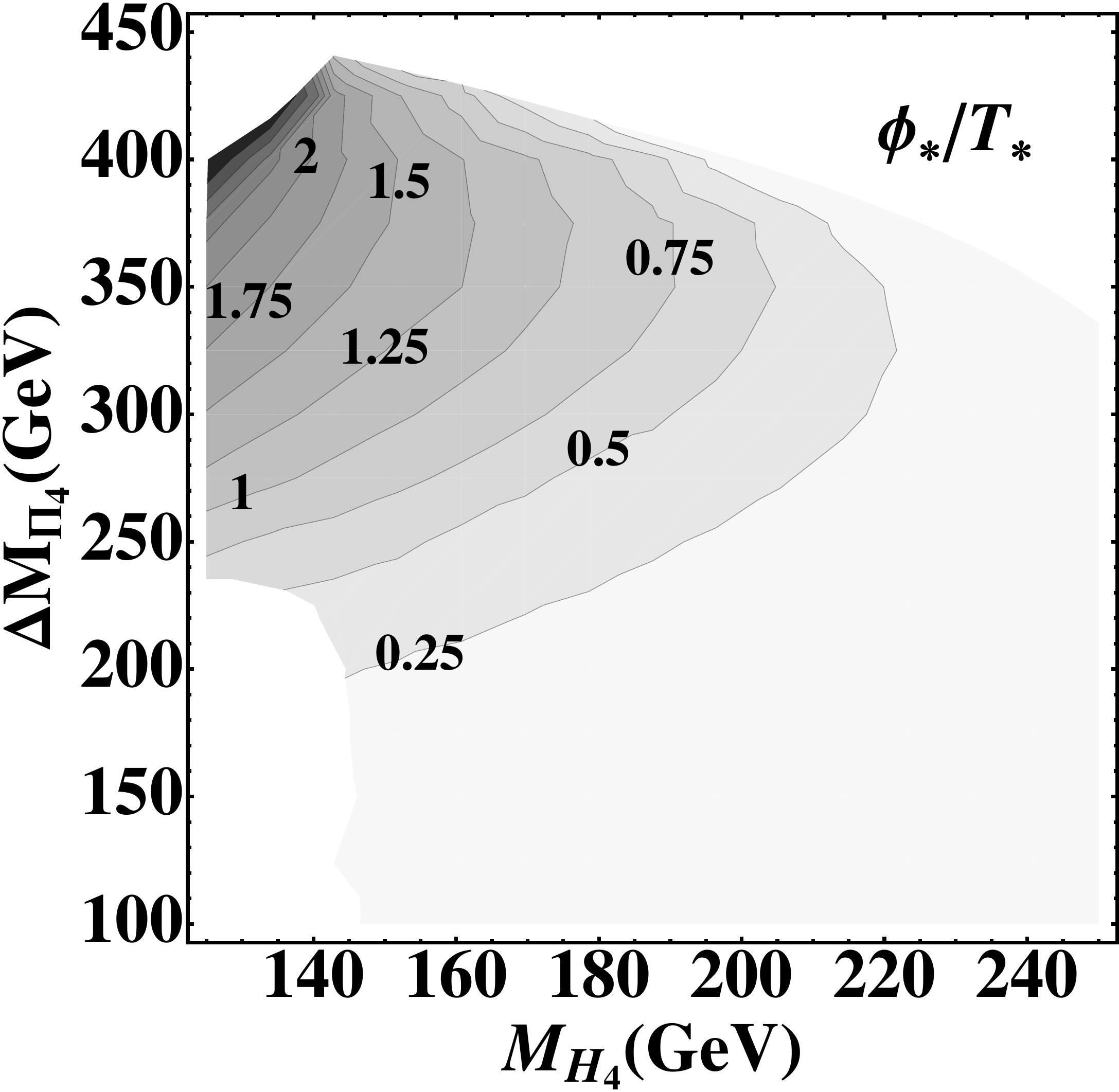}\hspace{0.1cm}\includegraphics[height=5.11cm,width=5.32cm]{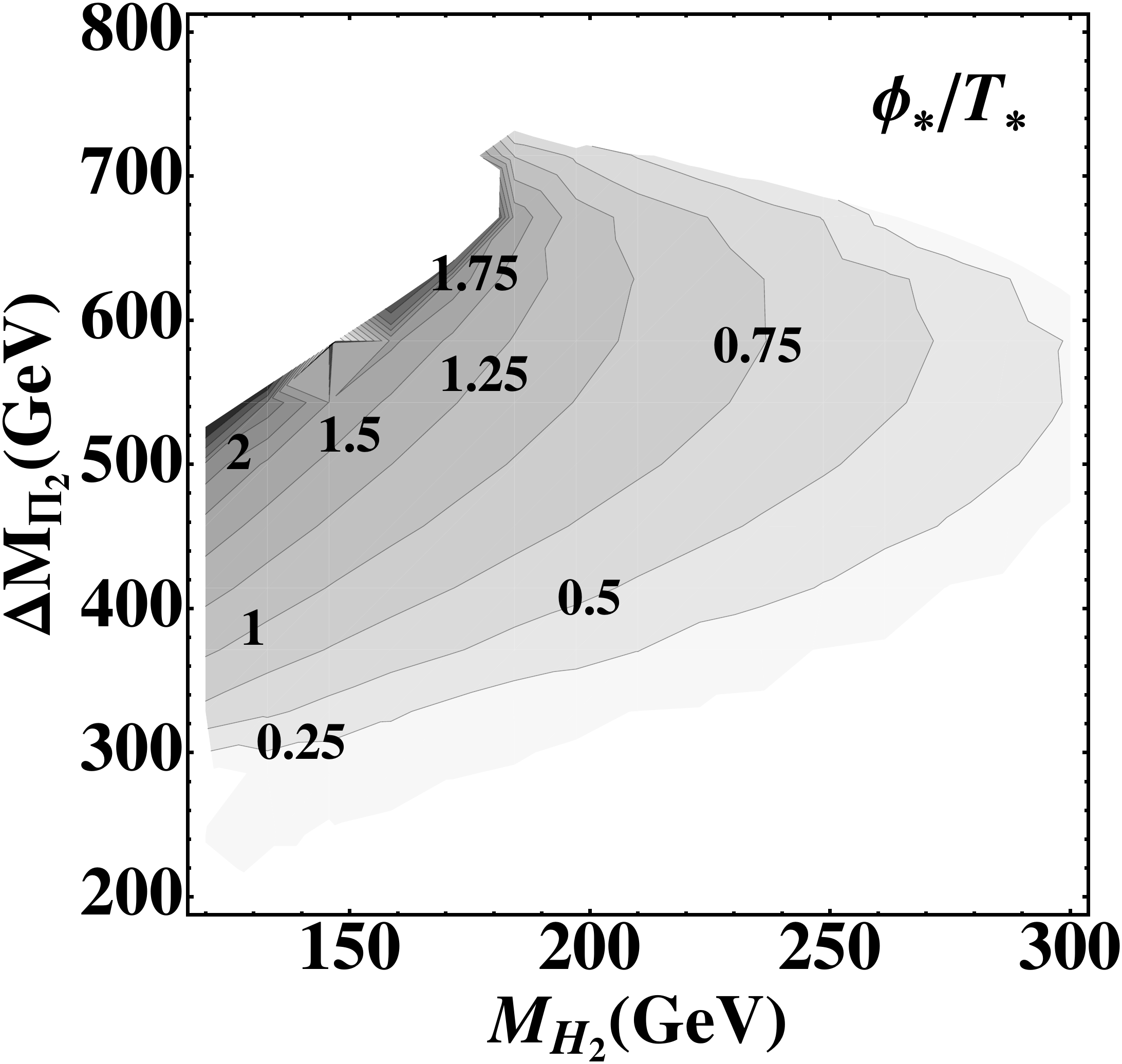}

\includegraphics[height=5.11cm,width=5.32cm]{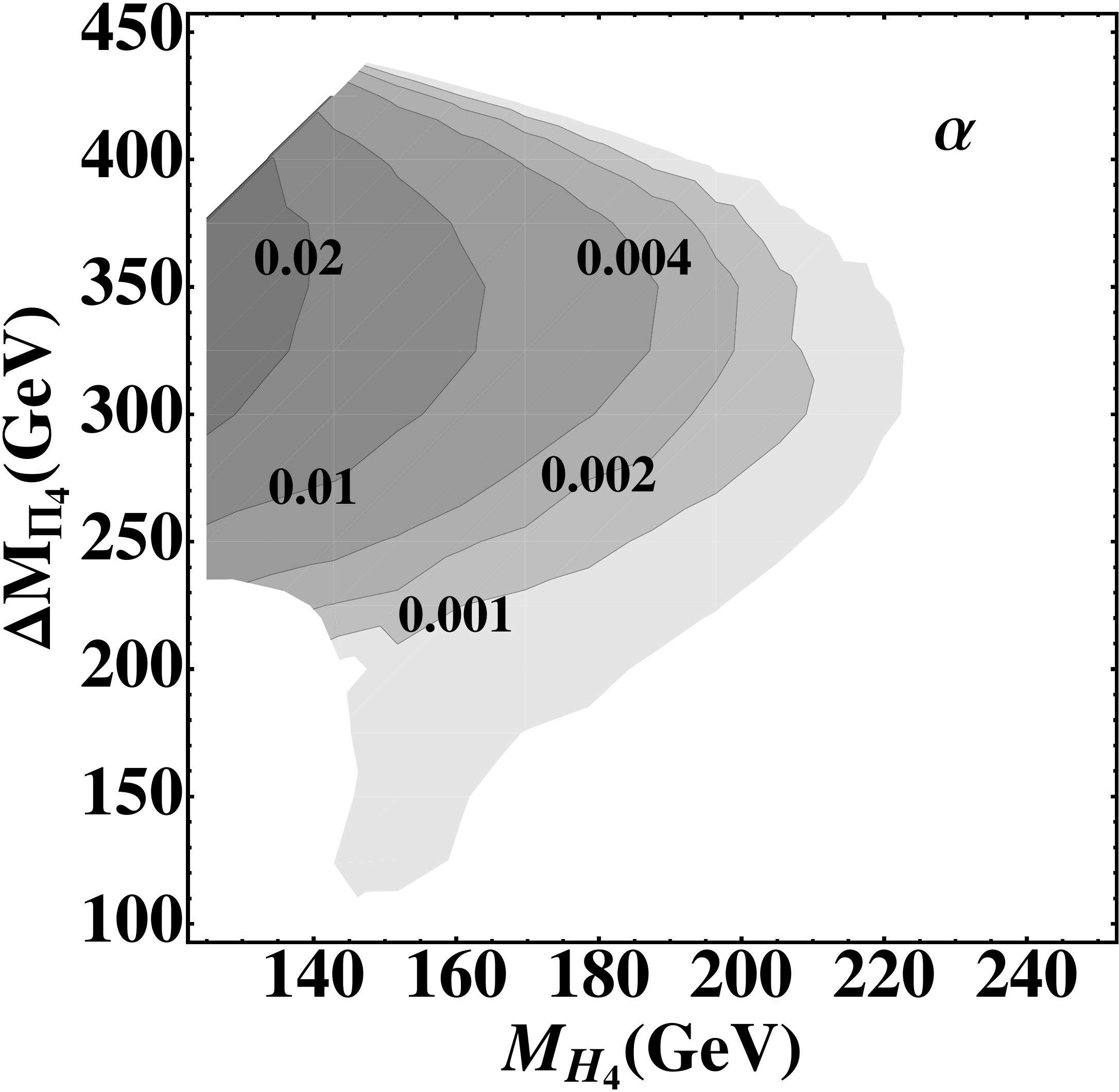}\hspace{0.1cm}\includegraphics[height=5.11cm,width=5.32cm]{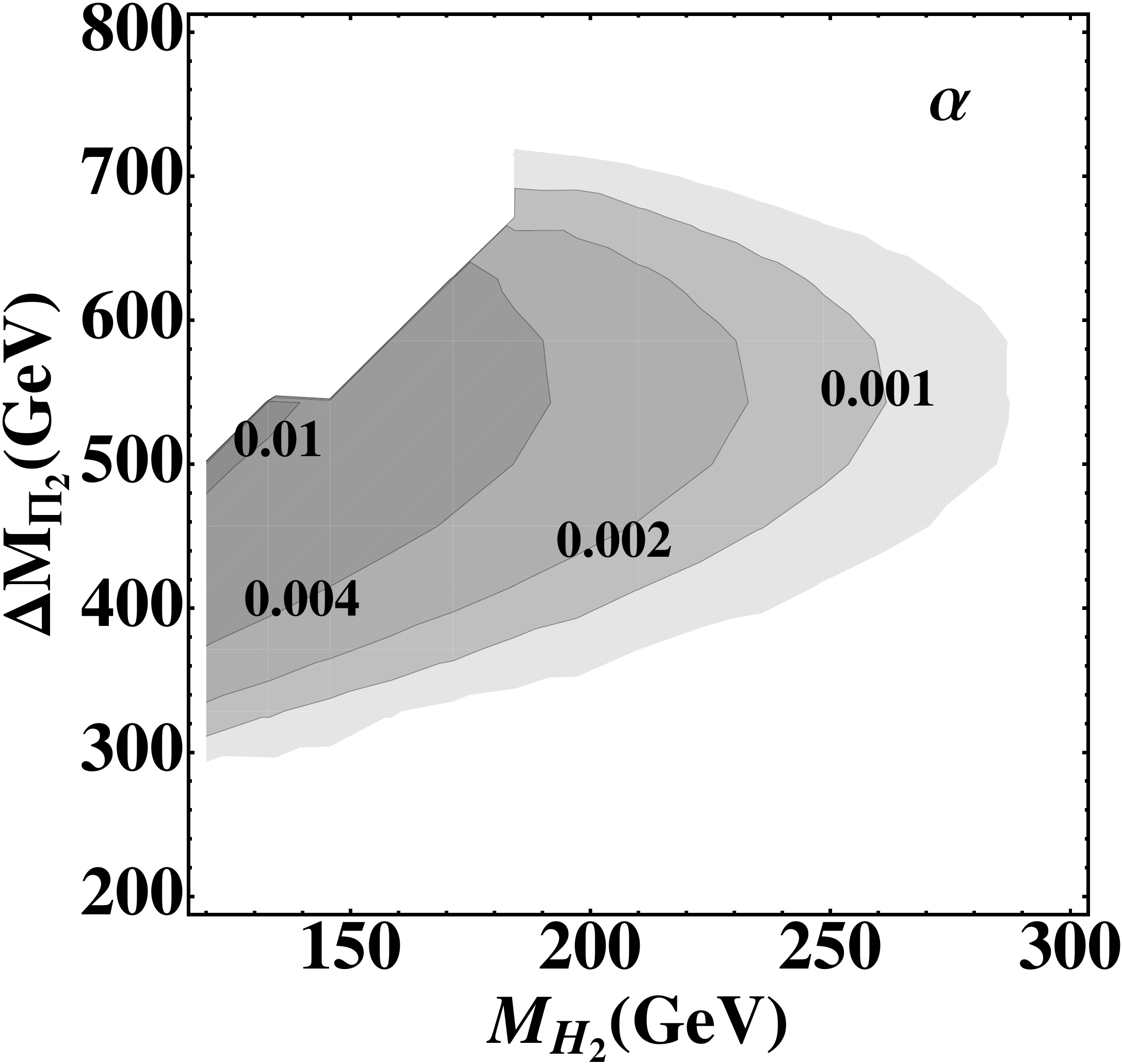}

\includegraphics[height=5.11cm,width=5.32cm]{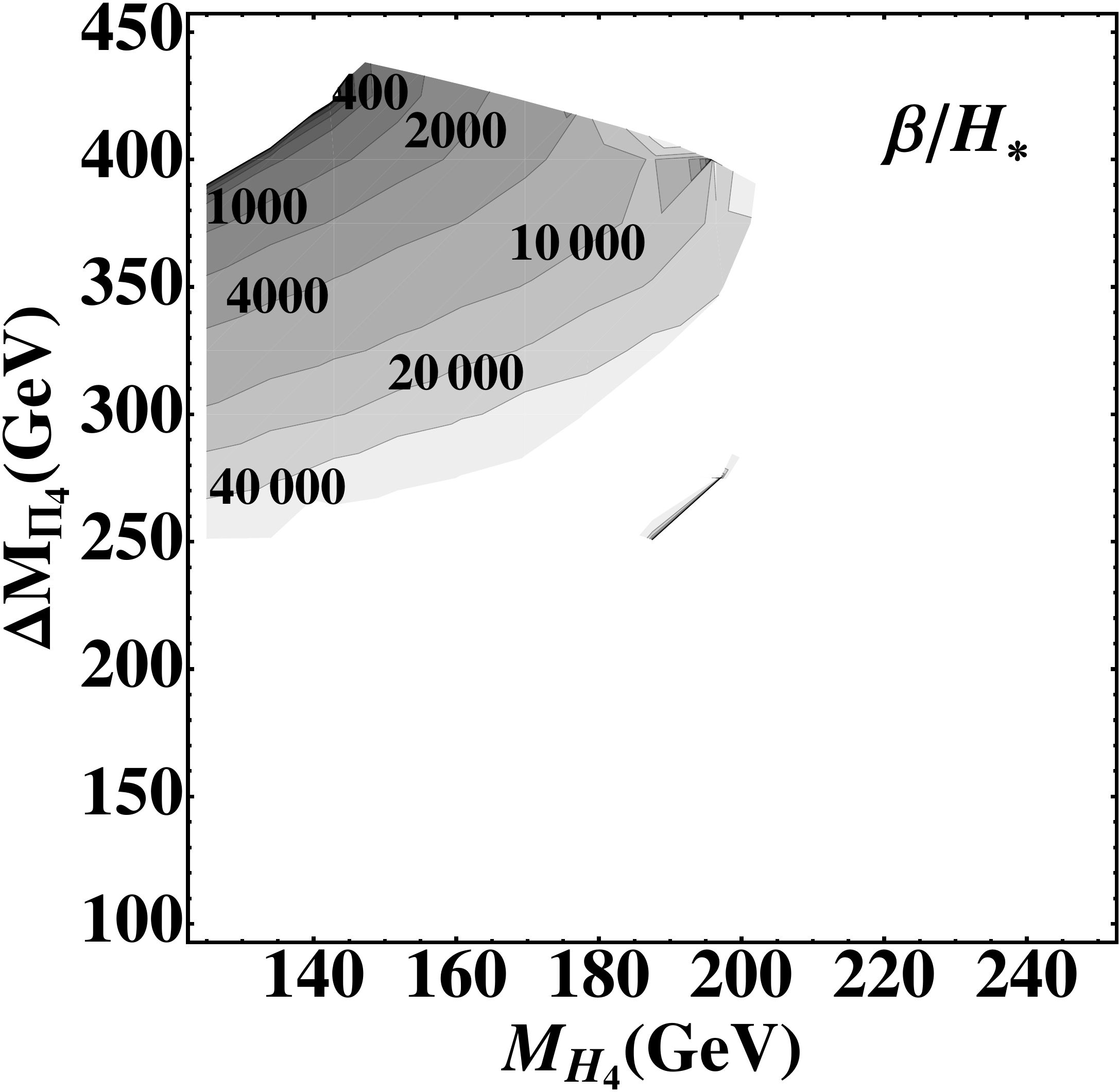}\hspace{0.1cm}\includegraphics[height=5.11cm,width=5.32cm]{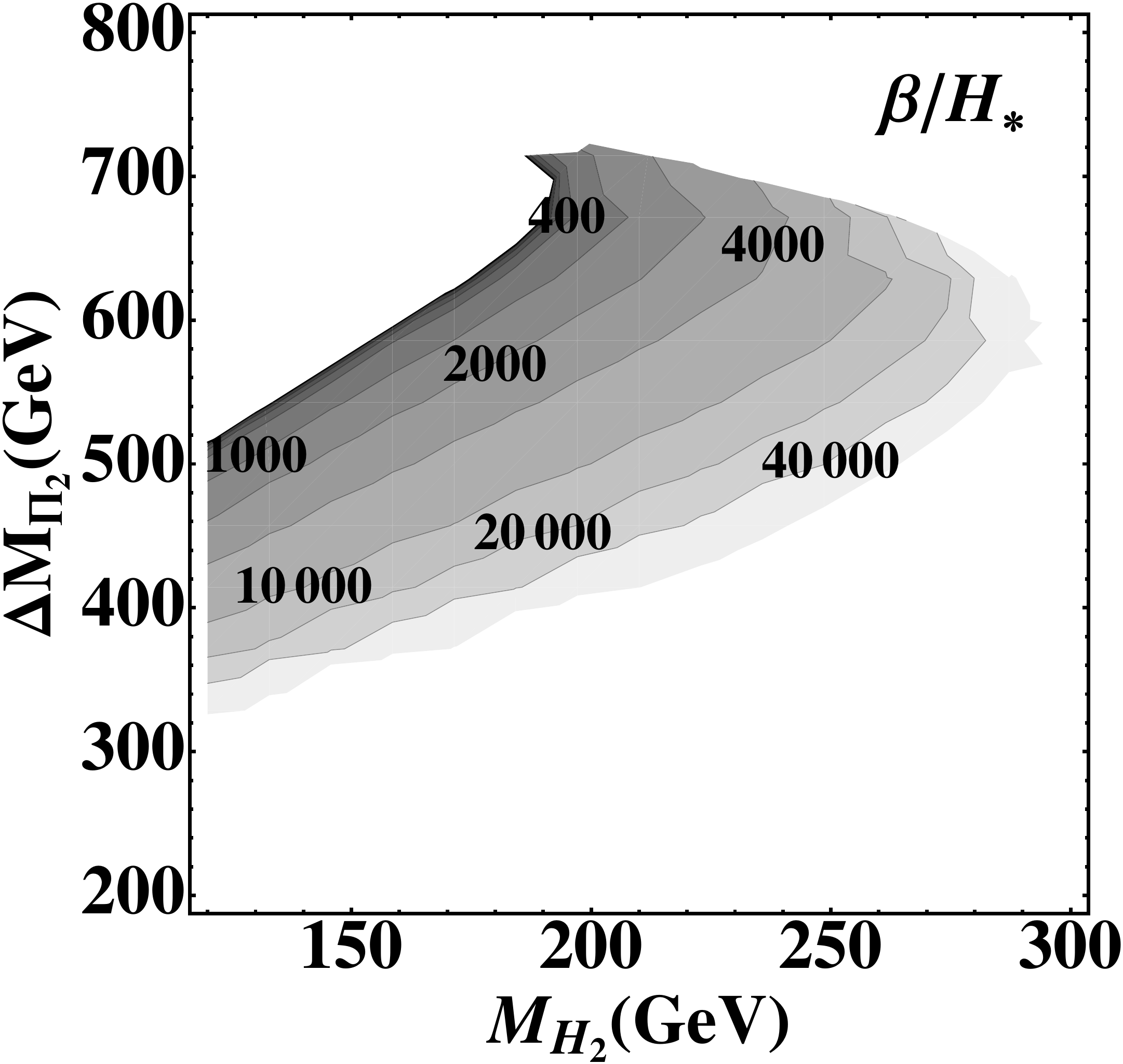}
}
\caption{The nucleation temperature $T_*$, the strength of the phase transition $\phi_*/T_*$, the parameter $\alpha$, and the parameter $\beta/H_*$ (from top to bottom) in the $M_H$-$\Delta M_{\Pi}$ plane for the ``4 transition'' (left) and for the ``2 transition'' (right) of UMT. 
We fixed ETC masses at 150~GeV and used $v_2=300$~GeV.}\label{fig:UMTres} \end{figure}

The effective Lagrangian of scalar particles in UMT includes several free parameters.
However, since we study only the case where the potentials related to the fundamental and adjoint techniquarks are decoupled, we have $\delta=0=\delta'$ and each of the transitions depends only on a certain subset of parameters. The  potential of the $SU(4)$ sector (fundamental quarks) is characterized by the Higgs mass $M_{H_4}$, the masses of the scalar partners of the Goldstone bosons, and the ETC masses. Since this sector breaks the electroweak symmetry, the zero-temperature VEV $v_4$ must equal the electroweak scale $246$~GeV. The dependence on the ETC masses is relatively weak, and we fix all of them to be 150~GeV.  For the (dynamical contribution to the) masses of the scalar partners of the Goldstone bosons we use
\bea
 \left(\Delta M_{\Pi_4}\right)^2 &\equiv& M_{\tilde \Pi_{UD}}^2 - m_{1,ETC}^2 = 2 \lambda^\prime_4 v_4^2 \ .
\eea
Similar parameters characterize the potential of the $SU(2)$ sector. In addition to the Higgs mass $M_{H_2}$ and
\bea
 \left(\Delta M_{\Pi_2}\right)^2 &\equiv& M_{\tilde \Pi_{\lambda\lambda}}^2 - m_{1,ETC}^2 = 4 \lambda^\prime_2 v_2^2
\eea
the zero-temperature value of the condensate $v_2$ is now a free parameter (while $v_4$ was fixed to 246~GeV). Notice also that since only the fundamental techniquarks are charged under the electroweak gauge symmetry, the $SU(2) \to SO(2)$ transition is independent of the standard model parameters.

\subsection{Results}

The results for UMT are shown in Fig.~\ref{fig:UMTres}.
The rows from top to bottom show the behavior of the nucleation temperature $T_*$, the ratio $\phi_*/T_*$ at the nucleation temperature, the parameter $\alpha$ characterizing the produced latent heat, and the parameter $\beta/H_*$ characterizing the rapidness of the transition, respectively. The left-hand plots give the parameters for the $SU(4) \to Sp(4)$ transition (the one coupled to the electroweak), while the right-hand plots are for the $SU(2) \to SO(2)$ transition.

In general the plots are very similar as in MWT above. The main difference is that both of the transitions here are weaker than in MWT: in particular, the crucial parameter $\alpha \lesssim 0.02$. Therefore, it seems unlikely that detectable GWs could be produced in the UMT model. 
The maximal value of the $\alpha$ parameter can be slightly enhanced by optimizing the choice for the ETC masses, possibly by adding new ETC operators, and by increasing the value of $v_2$ (which was fixed to 300~GeV above in Fig.~\ref{fig:UMTres}).

\begin{figure}[!tbp]
\begin{center}
\includegraphics[width=0.7\linewidth]{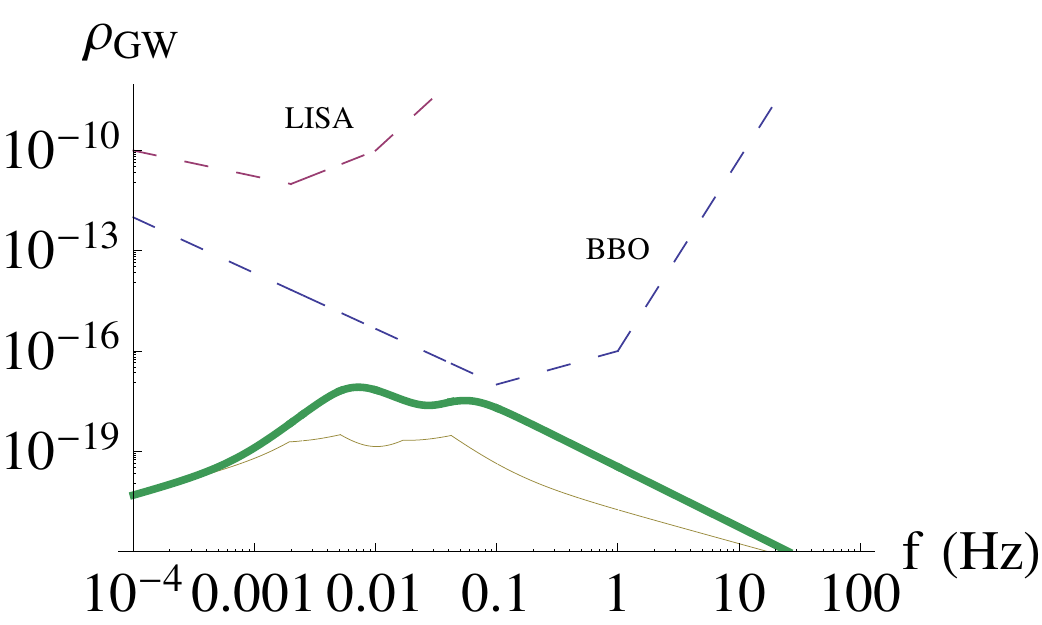}
\caption{The density of produced GWs $\rho_{\text{GW}}=\Omega_{\text{GW}} \text{h}^2$ in the case of the UMT as a function of the frequency (in Hz). Dashed lines represent the expected sensitivity of LISA and BBO, while solid lines represent the gravitational spectrum of bubble collision and turbulence combined. The thin and thick solid lines correspond to the gravitational spectrum with bubble collisions given, respectively, by Eqs.~(\ref{col2}),~and~(\ref{col3}). The spectrum from the $\sigma_4$ (electroweak) transition peaks at roughly $f=0.05$~Hz, while the spectrum from the $\sigma_2$ transition peaks around $f=0.005$~Hz. The values of the parameters are given in the text.} \label{fig:gw_uwt}
\end{center}
\end{figure}

Despite the weakness of the transition, there is however a very intriguing scenario. UMT admits successive phase transitions that can be of first order and occur at different temperatures. This in principle means that the gravitational spectra of the two (or more in general) phase transitions can peak at well separated frequencies. If the parameters $\alpha$ and $\beta$ are such that the spectrum of one phase transition does not completely cover the one of the second, then multiple peaks can be potentially seen. Such a case is depicted in Fig.~\ref{fig:gw_uwt}. In this case, one phase transition, i.e., the one associated to $\sigma_4$ (the one breaking the electroweak symmetry), has $\alpha=0.025$, $\beta/H_*=3000$, and $T_*=120$ GeV, while the second phase transition associated to $\sigma_2$ has $\alpha=0.005$, $\beta/H_*=200$, and $T_*=200$ GeV.
This set of values can be deduced from the effective potential with $M_{H_4} \simeq 130$~GeV and $\Delta M_{\Pi_4} \simeq 350$~GeV for the ``4 sector'' and with $M_{H_2} \simeq 180$~GeV and $\Delta M_{\Pi_4} \simeq 635$~GeV for the ``2 sector.'' The ETC masses were 150~GeV and $v_2=300$~GeV as above. As seen from Fig.~\ref{fig:UMTres} this is a rather optimal scenario with a low Higgs mass $M_{H_4}$ and parameters for the $\sigma_2$ transition rather near the critical line. 
In Fig.~\ref{fig:gw_uwt} we see the existence of multiple peaks that span almost two orders of magnitude in frequency (from $10^{-3}$ to $10^{-1}$). The peaks at higher frequency are due to the transition in the $\sigma_4$ sector, which also breaks the electroweak symmetry, while the lower peaks are produced by the $\sigma_2$ transition. Unfortunately as we see, the spectrum lies below the expected sensitivity of BBO. As mentioned above, the spectrum can be enhanced by choosing optimal values for the $ETC$ masses and the vacuum expectation value $v_2$. However, in all the configurations which we have checked, this does not change the conclusion: the nontrivial structure of the spectrum is hardly visible at BBO. Nevertheless, this is an interesting case for two reasons (apart from hoping for a better sensitivity of BBO). First, one can consider an underlying theory which has a larger number flavors than UMT. The presence of extra fermions in the theory can strengthen the phase transitions. In such a case, multiple peaks can be above the BBO sensitivity. Second, the estimation of GWs from first order phase transitions is far from conclusive. For example, apart from bubble collision and pure turbulence as we considered here, there is a possibility of producing GWs via primordial magnetic fields~\cite{Caprini:2001nb,Kahniashvili:2008er,Kahniashvili:2008pe}, with potentially larger amplitudes.

\section{Conclusion}
In this paper we studied the production of GWs from first order phase transitions of theories that dynamically break the electroweak symmetry. Although our setup is general, we looked, in particular, at two different models, i.e., the Minimal Walking Technicolor, and the Ultra Minimal Technicolor, that have been studied extensively. For MWT we found that there is parameter space, at the effective Lagrangian level,  for a sufficiently strong first order phase transition that produces GWs detectable at BBO. In this case, however, the phase transition is so strong that if a baryogenesis mechanism takes place, sphalerons would not be able to wash out the produced asymmetry. First principle lattice simulations will be able to disentangle, in the near future, the order of the phase transition of the underlying gauge theory. We should also stress that the low energy effective theory used here for the MWT Lagrangian can also describe the low energy effective theory for an $SO(4)$ gauge theory with 2 Dirac fermions in the fundamental representation of the gauge group. The latter gauge theory has two advantages over the traditional $SU(2)$ gauge theory with fermions in the adjoint representation. It is not expected to be conformal \cite{Sannino:2009aw} and does not feature technigluons-techniquark bound states \cite{Frandsen:2009mi} with potentially dangerous fractionally charged states. 
We have also discovered that, for reasonable values of the parameters, UMT seems not to be able to provide a very strong first order phase transition. However, UMT undergoes successive phase transitions, which can produce a gravitational spectrum of multiple peaks spanning two orders of magnitude in the frequency. This provides a very characteristic signal that can differentiate strongly coupled theories with multiple first order phase transitions.

\acknowledgments
We would like to thank Paul Hoyer, Thomas Konstandin, and Thomas A. Ryttov for useful discussions. The work of MJ was partially supported by the Villum Kann Rasmussen foundation.

\newpage
\appendix

\section{The effective Potential}
\label{app:effpot}

The effective potential is obtained by adding to the tree-level potential $V^{(0)}$ the one-loop correction $V^{(1)}$
\begin{eqnarray}
 V(\sigma) = V^{(0)}(\sigma) + V^{(1)}_{T=0}(\sigma) + V^{(1)}_{T}(\sigma,T) \ .
\end{eqnarray}
For brevity, we denote by $\sigma$ the expectation values of the Higgs field(s) that characterize the 
techniquark condensate(s). For MWT we have a single condensate $\sigma$, while for UMT we identify $\sigma =\{\sigma_4,\sigma_2\}$ where $\sigma_i$ refer to the condensates of the two sectors in UMT.
The standard zero-temperature one-loop contribution to the potential reads
\begin{eqnarray} \label{VT0}
V^{(1)}_{T=0} = \frac{1}{64\pi^2} \sum_{i} \bar{n}_i\, f_{i}(M_i(\sigma)) \ ,
\end{eqnarray}
where the index $i$ runs over all of the mass eigenstates and $\bar{n}_i$ is the multiplicity
factor for a given scalar particle $n_b$, while for Dirac fermions it is $-4$
times the multiplicity factor of the specific fermion $n_f$. The function $f_i$ is
\begin{equation} \label{fdef}
f_i = M^4_i(\sigma) \left[\log\frac{M^2_{i}(\sigma)}{M^2_i(v)}  - \frac{3}{2}\right] + 2M^2_i(\sigma) \, M^2_i(v)  \ ,
\end{equation}
where $M^2_i(\sigma)$ is the background dependent mass term of the
$i$th particle and $\sigma_{T=0} = v$.

The one-loop, ring-improved, correction can be divided into fermionic, scalar, and vector contributions,
\begin{eqnarray}
V_T^{(1)} = {V_T^{(1)}}_{\rm f}+ {V_T^{(1)}}_{\rm b}+ {V_T^{(1)}}_{\rm gauge}\ .
\end{eqnarray}
We use the extrapolation method introduced in \cite{Cline:1996mga} for evaluating the one-loop correction. At high temperatures we expand in $M_i/T$, which gives for the fermions
\begin{eqnarray} \label{VTf}
{V_T^{(1)}}_{\rm f,h}(N) &=& 2\frac{T^2}{24} \sum_{f} n_{f} M_{f}^2(\sigma) +\frac{1}{16 \pi^2}\sum_fn_{f} M_{f}^4(\sigma)\left[\log\frac{M_{f}^2(\sigma)}{T^2}-c_f\right] \nonumber\\
&& - 2\sum_{f} n_{f} M_{f}^2(\sigma)T^2 \sum_{l=2}^N\left(\frac{-M_{f}^2(\sigma)}{4\pi^2T^2}\right)^l \frac{(2l-3)!!\zeta(2l-1)}{(2l)!!(l+1)}\left(2^{2l-1}-1\right) \ ,
\end{eqnarray}
where $c_f \simeq 2.63505$.
For the bosons (including the electroweak gauge bosons, for which $n_b=3$) we write instead
\begin{eqnarray} \label{VTb}
{V_T^{(1)}}_{\rm b,h}(N) &=&\frac{T^2}{24} \sum_{b} n_{b} M_{b}^2(\sigma)  - \frac{T}{12\pi} \sum_b\,n_b \,M_{b}^3(\sigma,T) \nonumber\\ &&-\frac{1}{64 \pi^2}\sum_bn_{b} M_{b}^4(\sigma)\left[\log\frac{M_{b}^2(\sigma)}{T^2}-c_b\right] \nonumber\\
 && +\sum_{b} n_{b} \frac{M_{b}^2(\sigma)T^2}{2} \sum_{l=2}^N\left(\frac{-M_{b}^2(\sigma)}{4\pi^2T^2}\right)^l \frac{(2l-3)!!\zeta(2l-1)}{(2l)!!(l+1)} \ ,
\end{eqnarray}
where $c_b\simeq 5.40762 $.
At low temperatures we use for both bosons and fermions the asymptotic expansion of the one-loop correction
\be
 {V_T^{(1)}}_{\rm l}(N) = - \sum_i n_i e^{-M_i(\sigma)/T} \left(\frac{M_i(\sigma)T}{2\pi}\right)^{3/2} \sum_{l=0}^N \frac{1}{2^l l!}\frac{\Gamma(5/2+l)}{\Gamma(5/2-l)}\left(\frac{T}{M_i(\sigma)}\right)^l \ .
\ee

The extrapolated one-loop correction reads for each fermion
\be
 {V_T^{(1)}}_{\rm f} = \Theta\left(x_f- \frac{M_f(\sigma)^2}{T^2} \right) {V_T^{(1)}}_{\rm f,h}(N=4) + 4\Theta\left(\frac{M_f(\sigma)^2}{T^2}-x_f \right)\left({V_T^{(1)}}_{\rm l}(N=3)-\delta_f\right) \ ,
\ee
where $\Theta$ is the step function. The parameter $x_f \simeq 2.21605$ and the small correction $\delta_f \simeq -7.90454\times 10^{-4}$ were fixed by requiring the function to be continuous and differentiable with respect to $M_f$ at $M_f^2/T^2=x_f$. For the scalars we resum the contribution of the ring diagrams. Following Arnold and Espinosa
\cite{Arnold:1992rz} we write
\bea
 {V_T^{(1)}}_{\rm b} & = & \Theta\left(x_b- \frac{M_b(\sigma)^2}{T^2} \right) {V_T^{(1)}}_{\rm b,h}(N=3) + \Theta\left(\frac{M_b(\sigma)^2}{T^2}-x_b \right)\left({V_T^{(1)}}_{\rm l}(N=3)-\delta_b\right) \nonumber \\
                     &   &  + \frac{n_b T}{12 \pi} \left(M_b^3(\sigma)-M_b^3(\sigma,T)\right) \ ,
\eea
where  $x_b \simeq 9.47134$, $\delta_b \simeq 3.1931 \times 10^{-4}$, and $M_b(\sigma,T)$ is the thermal mass which follows from the tree-level plus one-loop thermal contribution to the potential. For the gauge bosons we set $n_b=3$:
\bea
 {V_T^{(1)}}_{\rm gb} & = & \Theta\left(x_b- \frac{M_{gb}(\sigma)^2}{T^2} \right) {V_T^{(1)}}_{\rm b,h}(N=3) + \Theta\left(\frac{M_{gb}(\sigma)^2}{T^2}-x_b \right)\left({V_T^{(1)}}_{\rm l}(N=3)-\delta_b\right) \nonumber \\
                     &   &  + \frac{T}{12 \pi} \left(M_{gb}^3(\sigma)-M_{L,gb}^3(\sigma,T)\right) \ .
\eea
Here $M_{L,gb}(\sigma,T)$  is the longitudinal mass of a given gauge boson and we have $M_{L,gb}(\sigma,T=0) = M_{gb}(\sigma)$ while the transverse mass receives only a suppressed temperature dependent correction which we have neglected.

For a more complete presentation with explicit expressions for the masses see \cite{Cline:2008hr,Jarvinen:2009pk}.


\end{document}